\documentclass[review]{elsarticle}

\usepackage[hyphens]{url}
\usepackage{lineno,hyperref}
\usepackage{graphicx,pstricks}
\usepackage{graphics}
\usepackage{epsfig}
\usepackage{amssymb}
\usepackage{amsmath, amsthm}
\usepackage{amsfonts}
\usepackage{algorithm}
\usepackage{tabu}
\usepackage{multirow}
\usepackage{subfig}
\usepackage{caption}

\begin{document}

\begin{frontmatter}

\title{A Principled Approach to Design Using High Fidelity Fluid-Structure Interaction Simulations}

\author[cee]{Wensi Wu\corref{correspondingauthor}}
\author[cee]{Christophe Bonneville\corref{correspondingauthor}}
\author[cee,cam]{Christopher Earls}

\cortext[correspondingauthor]{Equal contribution}
\address[cee]{School of Civil \& Environmental Engineering, Cornell University, Ithaca, NY 14850, United States}
\address[cam]{Center for Applied Mathematics, Cornell University, Ithaca, NY 14850, United States}

\begin{abstract}
A high fidelity fluid-structure interaction simulation may require many days to run, on hundreds of cores. This poses a serious burden, both in terms of time and economic considerations, when repetitions of such simulations may be required (\textit{e.g.} for the purpose of design optimization). In this paper we present strategies based on (constrained) \textit{Bayesian optimization (BO)} to alleviate this burden. BO is a numerical optimization technique based on Gaussian processes (GP) that is able to efficiently (with minimal calls to the expensive FSI models) converge towards some globally optimal design, as gauged using a \textit{black box} objective function. In this study we present a principled design evolution that moves from FSI model verification, through a series of \textit{Bridge Simulations} (bringing the verification case incrementally closer to the application), in order that we may identify material properties for an \textit{underwater, unmanned, autonomous vehicle (UUAV)} sail plane. We are able to achieve fast convergence towards an optimal design, using a small number of FSI simulations (a dozen at most), even when selecting over several design parameters, and while respecting optimization constraints.
\end{abstract}

\begin{keyword}
Bayesian Optimization \sep Fluid-Structure Interaction \sep Gaussian Process \sep Machine Learning \sep Design Optimization
\end{keyword}

\end{frontmatter}


\section{Introduction}
\noindent Fluid-structure interaction (FSI) analyses have been successfully employed in increasing numbers of engineering application areas over the past decades. A few examples of such efforts include FSI modeling of heart valves \cite{vigmostad2020}, prediction of aerodynamics flutter around an AGARD 445.6 wing \cite{kamakoti2004},  parachute inflation simulation in turbulent supersonic flows \cite{huang2018}, and slamming-induced structural response on a large container ship \cite{tuitman2009}. While the idea of effectively employing FSI analyses in design is attractive because of the enhanced insight it offers (\emph{i.e.} enabling study of the interplay of a solid deforming due to forcing from a fluid medium), correctly capturing the physical response of a multi-physics system, where each sub-system is described using a distinct set of governing equations, is a nontrivial endeavor. Particularly, in a partitioned FSI approach (\emph{i.e.} where the analysis context is such that the fluid and structural systems are considered using separate numerical descriptions, with external coupling strategies imposed to ensure satisfaction of salient transmission conditions along the fluid-structure interface) \cite{Matthies2003, Matthies2006, degroote2010_2}, a very fine temporal and spatial discretization is usually required to properly resolve the flow fields, and to correctly reproduce the physical mechanisms along the fluid-structure interface. In addition, in order to mitigate instabilities arising from the presence of the so-called \textit{added mass effects} (\emph{i.e.} the two-way influence of the adjacent fluid mass contacting the structure, which is especially pronounced in the context of incompressible fluids) \cite{Causin2005, Forster2006, Forster2007}, implicit coupling strategies are required to properly satisfy the conservation equations along the FSI interface. Hence, performing FSI analyses is rather computationally expensive: a serious problem when such simulations are required within the design cycle of some engineering artifact.\\
\newline
This computational cost often arises as an obstacle for optimizing structural designs subject to heavy fluid loads. Indeed, optimizing design parameters through random search, with the need to evaluate the design candidate with FSI analysis, at each new design point, would be very costly, both in time and in computational resources. Furthermore, trying random guesses offers very little hope for effectively finding the optimal candidate, especially as the number of design parameters increases. Standard numerical optimization methods (\textit{e.g.} gradient-based methods, \textit{etc.}) are typically not applicable in the present case because they require convexity in the design objective function, and further, usually require many objective function evaluations. Consequently in this paper, we present design optimization strategies based on Bayesian optimization (BO): a powerful optimization method that is well suited for optimizing black box objective functions that are potentially noisy. In addition, BO is typically able to achieve quick convergence to an admissible candidate with minimal objective function evaluations. BO has been successfully applied to multiple fields of science and engineering, such as material design \cite{zhang2019bayesian, Frazier_2015}, aerospace \cite{inproceedings}, biology \cite{ulmasov2015bayesian}, drug discovery \cite{8539993}, and hyperparameter tuning in data sciences \cite{8791696}. In the context of FSI, we particularly focus on BO with inequality constraints \cite{10.5555/3044805.3044997}, which allows to handle more realistic design problems (\textit{e.g.} if we want to reduce the drag of a hydrodynamic profile, we might want to do so while maintaining a minimal lift, maximum admissible structural deformation, \textit{etc.})\\
\newline
In this paper, We are proposing a principled approach to design when high-fidelity FSI simulations are required as part of the design loop. We begin with the identification of a verification problem that is ``close'' within the design space of our ultimate application (in our case, the Turek \& Hron FSI3 benchmark \cite{turek2006} fits our needs). We subsequently incrementally extend the verification context to bring it ever closer to our application, using what we call \textit{Bridge Simulations}. We then apply our, now trusted, FSI simulation capability to treat the case of composite material design for use in the sail plane of a UUAV, as a demonstration. If, when extending the proposed approach to other problems, the application in question does not have a suitable benchmark verification case in the literature, then one must resort to experimental validation of a simplified version of the design context: this may be advisable in all cases, if feasible.\\
\newline
The outline of this manuscript is as follows: Section \ref{solver} provides details concerning the FSI solution framework used in the present work, along with some elements of Bayesian optimization theory;  FSI validation results, as well as bridging simulation results, are presented in Section \ref{Turek} and Section \ref{examples}; Section \ref{UUAV} provides a demonstration on how the proposed BO/FSI method can be applied to the composite design for a UUAV sail plane; and Section \ref{conclusion} concludes the work.\\

\section{Methods}\label{solver}
\begin{figure}[h]
\centering
\includegraphics[width=1\textwidth]{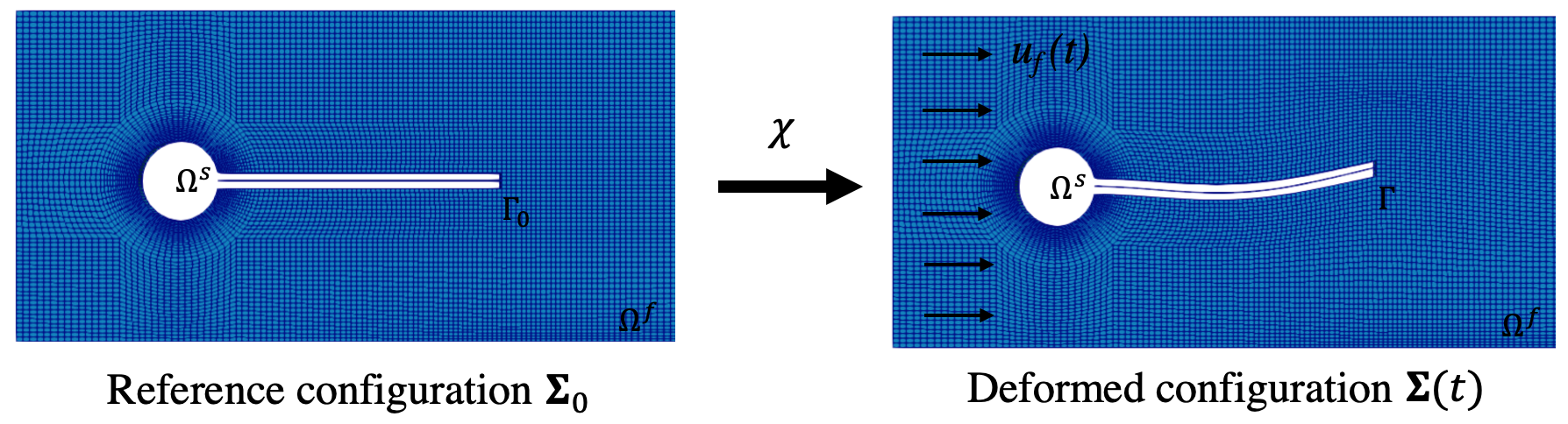}
\caption{A representative computational domain within an FSI system}\label{fsi_exm}
\end{figure}

\noindent A depiction of the computational domain for a representative fluid-structure interaction problem is presented in Figure \ref{fsi_exm}. Within the computational domain, a structural cantilever member, denoted as $\Omega_s$, is enclosed in a fluid domain, denoted as $\Omega_f$. The fluid domain (driven by an inflow defined by $u_f(t)$) interacts with the structural domain through the moving boundary $\Gamma$. Mesh deformation within the FSI system is governed by a mapping function $\chi$, which maps the mesh motion from its reference configuration $\Sigma_0$ to its deformed configuration $\Sigma(t)$. \\
\newline
The present work adopts a partitioned approach to FSI: relying to two open source analysis tools when treating the fluid and structural domains, respectively. The open source computational structural dynamics (CSD) finite element software used in this work is CU-BENs \cite{wu2020}, and the open source computational fluid dynamics (CFD) finite volume software, Open Field Operation and Manipulation (OpenFOAM 1.6-ext) \cite{jasak2007, weller1998, jasak1996, openfoam}, is used to treat the fluid domain. Information transfer between the CFD and CSD solvers is managed using a coupling library, denoted herein as the ``\textit{Coupler}'' \cite{Miller2010}.\\
\subsection{CU-BENs} \label{CUBENs}
\noindent CU-BENs \cite{wu2020} is an open source structural modeling finite element solver, written in C and developed at Cornell University. CU-BENs exploits high order structural elements for use in fully nonlinear (\emph{i.e.} both material and geometric contexts) transient dynamic analysis. The CU-BENs discrete Kirchhoff triangular (DKT) shell element is used in the present work, when modelling the structural responses within the FSI system. The Newmark implicit time integration scheme, augmented with generalized-$\alpha$ method \cite{chung1993}, for numerical stability purposes (artificial, numerical damping can improve solution behavior), is used for time integration within our structural  analyses.\\

\subsection{OpenFOAM 1.6-ext} \label{OpenFOAM}
\noindent OpenFOAM version 1.6-ext is a comprehensive CFD solver based on the finite volume method. In many ways, OpenFOAM functions as an object-oriented application programming interface (API) for C++. In this work, we employ OpenFOAM 1.6-ext, since it provides a FEM decomposition framework \cite{jasak2007, jasak2009} (also known as ``mini-element" method), along with a Laplacian-based mesh motion equation technique \cite{jasak2007, kassiotis2008, jasak2009}, to govern the mesh motion within the fluid domain. We found the finite element vertex-based mesh motion solver to produce superior numerical results, as compared with some of the other popular mesh motion methods (\emph{e.g.} the spring analogy \cite{Batina1989, blom2000} and its variants \cite{farhat1998_2, zhao2002, perot2003, degand2002}), as it guarantees boundedness in the Laplacian operator (the Laplace equation governs mesh vertex motion), even when the fluid cells approach a degenerate state \cite{Jasak2004, jasak2007}. This feature is vital for FSI simulations involving large structural displacements; since such structural responses could lead to considerable deformation and skewness of the attached fluid control volumes. \\
\subsection{Coupler} \label{Coupler}
\begin{figure}[h]
\centering
\includegraphics[width=1\textwidth]{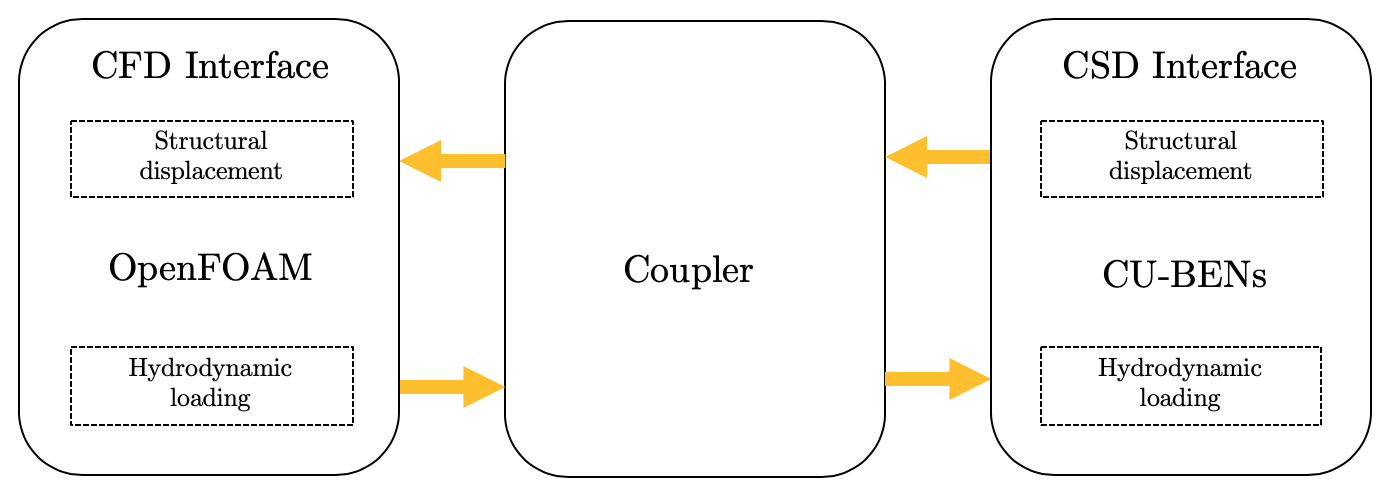}
\caption{Implicit partitioned FSI coupling framework with OpenFOAM as the fluid solver, CU-BENs as the structural solver, and a Coupler that facilitates all necessary communications between the two solvers to ensure the fluid-structure interface conforms to the transmission conditions}\label{coupling_framework}
\end{figure}

\begin{figure}[h]
\centering
\includegraphics[width=0.7\textwidth]{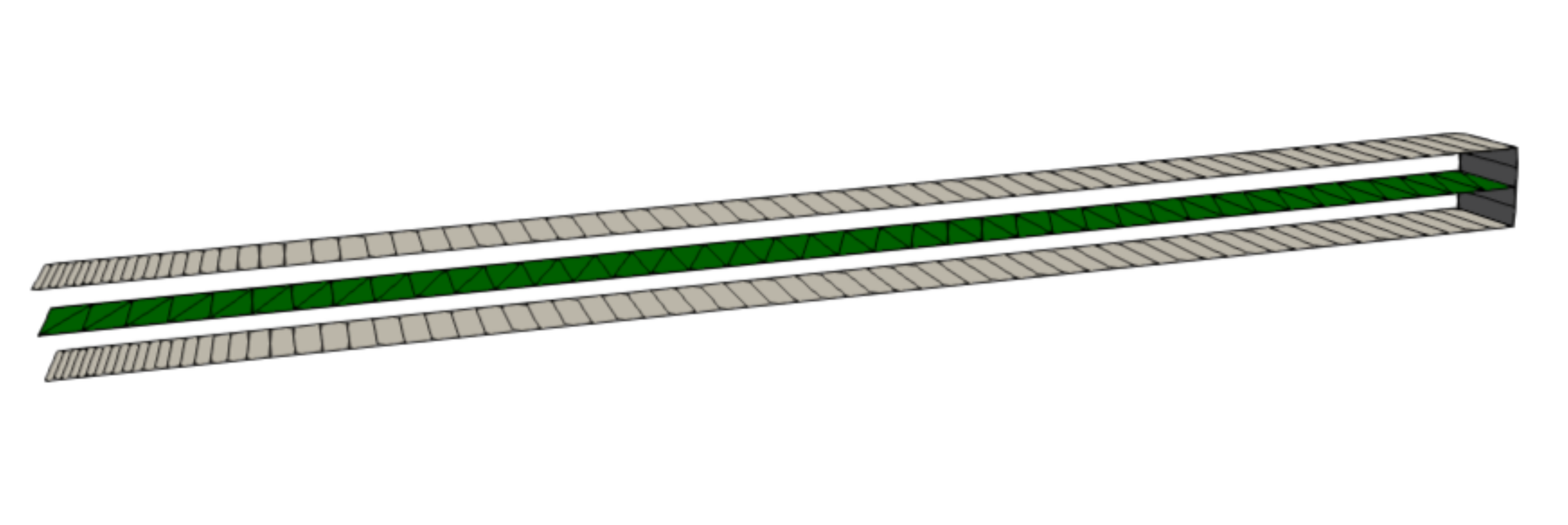}
\caption{Nonconforming mesh: grey represents the fluid boundary that interact with the structural elements, while green represents the structural DKT shell elements}\label{nonunimesh}
\end{figure}

\begin{figure}[h]
\centering
\includegraphics[width=0.25\textwidth]{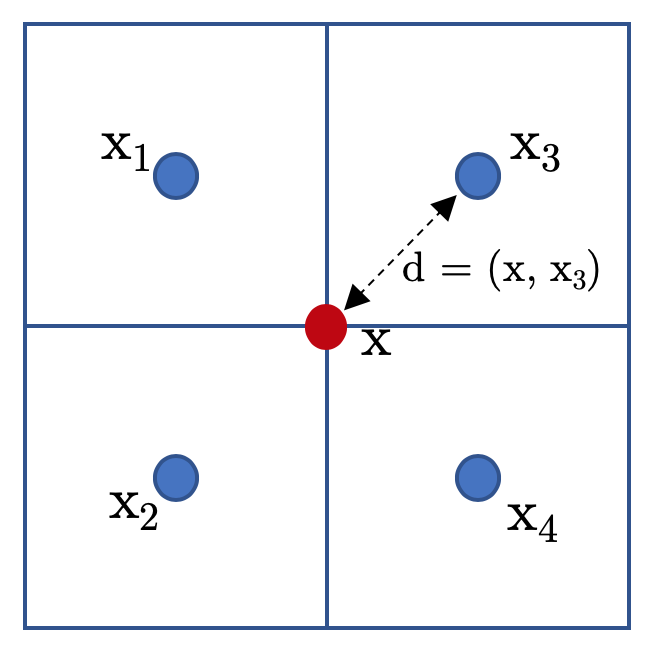}
\caption{Inverse distance weighting interpolation for converting fluid pressure at cell center (blue) to the point load (red)}\label{idw}
\end{figure}

\noindent The coupling library (the Coupler), initially developed by the Navel Surface Warfare Center Carderock Division (NSWCCD) \cite{Miller2010}, was modified for use in the present work. Due the extreme importance of the Coupler in FSI analyses, more detail is offered in what follows. In terms of functionality, the Coupler ensures that the salient transmission conditions are satisfied along the fluid-structure interface. Also, the Coupler manages data transfer between the CSD and CFD models. To effect communication with the Coupler, solver interface classes (\emph{i.e.} CFD interface and CSD interface) are written for OpenFOAM 1.6-ext and CU-BENs. Specifically, these C++ interface classes work with the respective CSD and CFD solvers to: prepare necessary data structures; adapt cross domain variables; and translate native input/output to a form that is compatible with the Coupler. The interface classes are subsequently translated to ``Python-ized" modules using the Python wrapper generator: SWIG. The resulting python files are then integrated into the Coupler. This allows the interface classes to communicate with the Coupler using pointers to memory; thus eliminating the need for writing and reading input/output files. A schematic representation of our implicit partitioned FSI solver coupling framework is presented in Figure \ref{coupling_framework}.\\
\newline
Within a given FSI simulation, it is the CFD solution that is most demanding, both in terms of time step size (time increment) and spatial discretization. The CSD solver can accommodate any time step size (due to its implicit time integration) and requires much coarser meshes (as a result of the high order elements). The orchestration of the FSI solutions, within each time step, follows the general scheme: \\
\begin{enumerate}
	\item OpenFOAM computes the hydrodynamic loading, given the structural deformation from the previous time step, using an \textit{Arbitrary Lagrangian Eulerian (ALE)} scheme.
	\item The CFD interface object takes in the resulting fluid fields and prepares necessary data for the Coupler.
	\item The Coupler interpolates the hydrodynamic loading, and subsequently maps the loading onto the structural mesh.
	\item CU-BENs then computes the corresponding structural deformation.
	\item The CSD interface class takes in the resulting FSI boundary deformation and prepares necessary data for the Coupler.
	\item The Coupler checks for convergence by monitoring out of balance displacement and/or pressures between the fluid and structural domains: comparing to a set of user specified tolerances. If the response along the interface converges, the simulation advances to the next time step. Otherwise, the predicted displacements are passed to OpenFOAM, to update the mesh configuration, and the process is repeated.
\end{enumerate}
\noindent Nonconforming mesh projection \cite{Farhat1998} is utilized to guide the grid-to-grid mesh mapping and updating along the fluid-structure interface. This technique allows for an optimal choice of mesh, for resolving the physics within the respective structural and fluid domains, as it provides flexibility for load and motion transferring on the FSI interface, when the mesh sizes in each subsystem are incongruous. Figure \ref{nonunimesh} provides a visual representation of nonconforming mesh in a FSI analysis. The structural domain is discretized with DKT shell elements; thus thickness is implied as a shell property (and not explicitly modeled). In such a context, the shell elements (green) are situated at the mid plane of the physical structure, and enclosed by the fluid boundary (grey). Orthogonal projection mapping is initiated at the beginning of a FSI simulation, to identify appropriate coupling pairs within the structural and fluid node sets; thus enabling mesh motion tracking throughout the analysis. It is typically the case in FSI analyses that the number of CFD nodes, occurring along the FSI interface, far exceeds the number of CSD nodes; thus leading to non-conforming mesh geometries that complicate data transfer in the FSI transmission conditions. The Coupler uses a standard approach when handling this case.\\
\newline
Within our Coupler, inverse distance weighting interpolation (IDW) \cite{shepard1968} is used to convert hydrodynamic pressure, located at the center of each fluid patch (one face of the finite volume), along the interface, into point loads, before they are projected onto the structural mesh. The IDW procedure used in the current work, employs the following relation:
\begin{equation}
F(x) = \sum_{i=1}^{n} \frac{\omega_i(x)F(x_i)}{\sum_{j=1}^{n}\omega_j(x)}, \quad \omega_i(x) = \frac{1}{d(x, x_i)^\alpha}
\end{equation}
where $x_i$ denotes the spatial coordinate of cell center $i$, $F(x_i)$ represents the fluid pressure at $x_i$, $n$ is the total number of nodal points used in the interpolation, $\omega_i$ stands the weighting factor, $d(x, x_i)$ corresponds to the distance between the node $x$ and cell center $x_i$, $\alpha$ is the distance-decay parameter, and $F(x)$ is the estimated point load at location $x$. Typically, $\alpha =2$ is used in the standard IDW. A simple representation of the fluid interface is shown in Figure \ref{idw} to help visualize the IDW interpolation process. \\

\noindent Time step advancement, and solution convergence on the fluid-structure interface, are controlled by the Coupler using an iterative method: the IQN-ILS iteration scheme \cite{degroote2009,degroote2010_2}. The IQN-ILS is a quasi-Newton root finding method, within which Newton-Raphson iterations are used, within a given time step, to solve for the interface displacement $\mathbf{R}(\mathbf{d}) = S(F(\mathbf{d})) =  0$. In the equation mentioned, $F()$ represents a CFD solver for a given displacement $\mathbf{d}$, while, $S()$ is a CSD solver for a given loading resulting from the CFD solver. The Newton-Raphson iteration formulation within each time step is shown in Equation \ref{2_7_14a} and \ref{2_7_14b}
\begin{equation} \label{2_7_14a}
\mathbf{R}^{k}+\mathbf{R'}^{k}\Delta\mathbf{d}^{k} = 0
\end{equation}
\begin{equation} \label{2_7_14b}
\mathbf{d}^{k+1} = \mathbf{d}^{k}+\Delta\mathbf{d}^{k}
\end{equation}
Combining Equation \ref{2_7_14a} and \ref{2_7_14b}, a new expression for $\mathbf{{d^{k+1}}}$ is derived
\begin{equation} \label{2_7_16}
\mathbf{d}^{k+1} = \mathbf{d}^{k}-(\frac{\mathbf{\delta R'}^{k}}{\mathbf{\delta d}})^{-1}\mathbf{R}^k,
\end{equation}
where $\mathbf{R'}^{k}$ represents the Jacobian of the interface displacement residual operator. Moreover, the residual vector $\mathbf{R^k}$, which is evaluated at $\mathbf{d}^k$, is described in Equation \ref{2_7_15}
\begin{equation} \label{2_7_15}
\mathbf{R}^k= S(F(\mathbf{d}^k))-\mathbf{d}^k= \mathbf{\tilde{d}}^{k+1}-\mathbf{d}^k.
\end{equation}
Herein, $\mathbf{\tilde{d}^{k+1}}$ represents the predicted FSI interface displacement at iteration $k+1$, whereas $\mathbf{{d^{k+1}}}$ is the corrected FSI interface displacement at iteration $k+1$.
The Newton-Raphson iteration step is repeated until $\mathbf{R}^k$ satisfies a user specified convergence criterion $\left\| \mathbf{R}^{k}\right\|\leq\epsilon$, where $\epsilon$ is a convergence tolerance (standard Euclidean norm, implied). \\
\newline
The procedure of IQN-ILS may seem easy enough when the exact Jacobian of $\mathbf{R}$ is known. However, in many cases, the Jacobians of the fluid and structural domains are not readily available; as in the case of coupling black box CFD/CSD pairs. In such cases, inverting $\mathbf{R'}^k$, to compute $\Delta \mathbf{d}^k$, is no longer a trivial task. Fortunately, we can circumvent the problem by constructing a set of basis vectors on $\mathbf{R}$ and $\mathbf{\tilde d}$, such that the solution to $(\frac{\mathbf{\delta R'}^{k}}{\mathbf{\delta d}})^{-1}\mathbf{R}^k$ can be approximated (\emph{i.e.} the inverse of the Jacobian does not need to be formed explicitly). A sketch of this procedure follows next.\\
\newline
When approximating the required Jacobian, a series of simple steps are used, such that their collective forms the required linearization within the IQN-ILS algorithm.
First, the difference of residual and interface displacement, from the previous iteration (denoted as superscript $k-1$) and the current iteration (denoted as superscripted $k$), are computed for iteration $i = 0, 1, 2, 3, ..., k-1$.
\begin{equation} \label{2_9_1}
\Delta\mathbf{R}^{k-1}=\mathbf{R}^{k}-\mathbf{R}^{k-1}
\end{equation}
\begin{equation} \label{2_9_2}
\Delta\mathbf{\tilde d}^{k+1} = \mathbf{\tilde d}^{k+1}-\mathbf{\tilde d}^{k}
\end{equation}
Each $\Delta\mathbf{R}^i$ corresponds to a $\Delta\mathbf{\tilde d}^{i+1}$, and these vectors are stored as columns within the matrices $V^k$ and $W^k$
\begin{equation}
V^k = \begin{bmatrix} \Delta\mathbf{R}^{k-1} & \Delta\mathbf{R}^{k-2} & ... & \Delta\mathbf{R}^1 & \Delta\mathbf{R}^0 \end{bmatrix}
\end{equation}
\begin{equation}
W^k = \begin{bmatrix} \Delta\mathbf{\tilde d}^{k} &  \Delta\mathbf{\tilde d}^{k-1} & ... &  \Delta\mathbf{\tilde d}^2 & \Delta\mathbf{\tilde d}^1 \end{bmatrix}
\end{equation}
where the number of columns, determined by the number of iterations, $q$, is limited by the degrees-of-freedom, $p$, on the FSI interface, such that the system is over-determined. In the event that the number of iterations exceeds the cardinality in degrees-of-freedom along the FSI interface, the rightmost column of $V^k$ must be truncated, to preserve the over-determined characteristic of the system: this allows the system to be solved as a least square problem \cite{golub2013}. \\
\newline
Subsequently, a set of orthogonal basis vectors is constructed, where the value of $\Delta\mathbf{R^k}$ is approximated as a linear combination of the known $\Delta\mathbf{R}^i$, as:
\begin{equation} \label {2_7_17a}
\Delta \mathbf{R^k} \approx \sum_{i=0}^{k-1} \alpha^k_i \Delta\mathbf{R}^i = V^k \alpha^k
\end{equation}
with $\alpha_i^k \in  \mathbb{R}^{q\times1}$. The matrix $V^k$ is then decomposed via economy QR decomposition, using Householder transformations
\begin{equation}\label {2_7_17b}
V^k=Q^kR^k
\end{equation}
into an orthogonal matrix $Q^k \in  \mathbb{R}^{p\times q}$, and an upper triangular matrix $R^k \in  \mathbb{R}^{q\times q}$. Substituting Equation \ref{2_7_17b} into Equation \ref{2_7_17a} yields
\begin{equation} \label {2_7_17c}
R^k\alpha^k=(Q^k)^T\Delta{\mathbf{R^k}},
\end{equation}
where the coefficient vector $\alpha^k$ in Equation \ref{2_7_17c} can be obtained by backward substitution. \\
\newline
Similarly, the corresponding $\Delta\mathbf{\tilde d^{k+1}}$ can also be calculated as a linear combination of the $\Delta\mathbf{\tilde d^i}$, from previous iterations,
\begin{equation} \label {2_7_18}
\mathbf{\Delta \tilde d^{k+1}} \approx \sum_{i=0}^{k-1} \alpha^k_i \mathbf{\Delta \tilde{d}}^i = W^k \alpha^k.
\end{equation}
From Equations \ref{2_7_14b}, \ref{2_7_15}, \ref{2_9_1} and \ref{2_9_2}, $\Delta \mathbf{R^k} = \Delta \mathbf{\tilde d^{k+1}}-\Delta \mathbf{d^k}$ is derived, this leads to the relation
\begin{equation} \label {2_7_19}
\mathbf{\Delta d^k} =  W^k \alpha^k - \Delta \mathbf{R^k}.
\end{equation}
Given that $\mathbf{\Delta d^k} =  -(\frac{\mathbf{\delta R'}^{k}}{\mathbf{\delta d}})^{-1}\mathbf{R}^k$ and $\Delta \mathbf{R^k}= -\mathbf{R}^k$ (assuming in an ideal case $R^{k+1}=0$), the quasi-Newton Raphson iteration step in Equation \ref{2_7_16} is then re-expressed as
\begin{equation} \label {2_7_19}
\mathbf{d}^{k+1} = \mathbf{d}^{k}+W^k \alpha^k + \mathbf{R^k}.
\end{equation}

\subsection{Bayesian Optimization}\label{botheory}
\noindent We now discuss the second critical piece within our FSI design ``puzzle'': Bayesian optimization. Bayesian optimization (BO) is a powerful global optimization framework that permits the minimization of complex, non-convex objective functions \cite{NIPS2012_4522, shahriari, frazier2018tutorial}. Unlike gradient based optimization methods, BO does not require calculation of gradients to the objective function, and typically only requires few evaluations of the objective function, in order to achieve satisfactory convergence towards the global minimum. Consequently, it is a method well suited for minimizing (noisy) black box functions that are expensive to evaluate (\textit{e.g.} experimental evaluations, computationally intensive evaluations, \textit{etc.}) The general optimization problem may be stated as follows:

\begin{equation}
    \label{optim}
    \text{find}\,\,\,\,\, \hat{f}=\min_{x\in\Omega}f(x)\,\,\,\,\,\text{s.t.}\,\,\,\,\,c_k(x)\leq \lambda_k \,\,\,\,\,(\,k\in[\![1,n_c]\!]\,)
\end{equation}

\noindent Where $f$ is the objective function, $\hat{f}$ its global minimum, $\Omega$ the domain of definition of $f$, and $c_k$ is a set of $n_c$ inequality constraints ($c_k$ may be a black box function that is evaluated at the same time as $f$) \cite{10.5555/3044805.3044997}.
\\\\
The main idea of BO is to approximate $f$ with a Bayesian regression surrogate that interpolates on a set of already known values of $f$. The Bayesian regression is vital, since it outputs ``confidence intervals'' on its prediction; thus providing useful hints concerning the veracity of our interpolations of the objective function. Additionally, this built-in \textit{uncertainty quantification} on the objective function allows for a principled formulation for the requisite \textit{acquisition function}; used in guiding our search for a minimum towards the most promising regions of the design space. Evaluating $f$ within the regions of design space that are specified by the acquisition function provides new data that are subsequently used to refine the Bayesian regression surrogate, and update the acquisition function prediction \cite{frazier2018tutorial}. Iterating this process usually yields convergence with only a small number of objective function evaluations.
\\\\
In the following subsections, we first provide more details about Gaussian processes (GP) \cite{10.5555/1756006.1953029}, which is the most widely used Bayesian regression method in BO. Then, we provide more insights concerning acquisition functions and constrained Bayesian optimization (cBO) \cite{10.5555/3044805.3044997}.

\subsubsection{Gaussian Process Regression}

In this brief introduction of GP regression, we employ the same notations as in \cite{gpshock}. Let us consider a machine learning training dataset made of $n$ input-output couples (\textit{training data}): $\mathcal{D}=(X,y)=(x_i,y_i)_{i\in[\![1,n]\!]}$. $x_i$ is an input vector (\textit{e.g.} the optimization/design parameters in the case of BO) and $y_i$ a collection of output scalars (\textit{e.g.} the corresponding value of the objective function in the case of BO). $y_i$ is assumed to be contaminated with Gaussian white noise (other noise models can be accommodated) and modeled as follows:

\begin{equation}
\label{erm}
\left\lbrace
\begin{array}{ll}
y_i=f(x_i)+\epsilon_i\\\\
\epsilon_i\sim\mathcal{N}(0,\sigma^2)\\
\end{array}\right.
\end{equation}

\noindent In machine learning (of which GP regression is a type), the main goal is to learn a representation of the unknown function $f$, given the available training data, in order to make prediction for new (yet unobserved) input points $x_*$ \cite{10.5555/1162264, hastie01statisticallearning, murphy2013machine}. Assumptions must be made about the functional form  of $f$, and in GP regression it is taken as a stochastic vector sampled from a multivariate Gaussian distribution \cite{10.5555/1756006.1953029} :
\begin{equation}\label{prior}
    f\sim\mathcal{GP}(0,k(X,X))
\end{equation}
The covariance matrix $k(X,X)$ depends on the input training data, and is computed using an arbitrary covariance kernel function. The kernel function, $k$, is a design choice that reflects prior knowledge one may have about the nature of $f$ (smoothness, linearity, \textit{etc.}) In BO, it is often hard to have any intuition about how the objective function behaves (especially in black-box optimization). A popular choice, that we will adopt in the following, is the $5/2$ Matérn kernel, described in Equation \ref{matern}. This kernel is well suited for handling both smooth and non-smooth functions \cite{frazier2018tutorial}.

\begin{equation}\label{matern}
    k(x_i,x_j;\theta)=\frac{\theta_1}{\Gamma(\nu)2^{\nu-1}}\bigg(\frac{\sqrt{2\nu}}{\theta_2}|\!|x_i-x_j|\!|\bigg)^\nu K_\nu\bigg(\frac{\sqrt{2\nu}}{\theta_2}|\!|x_i-x_j|\!|\bigg)
\end{equation}

\noindent where $\nu=5/2$, $\Gamma$ is the standard gamma function, $K_\nu$ is the modified Bessel function of the $2^\text{nd}$ kind, and $\theta=(\theta_1,\theta_2)$ is a set of tunable hyperparameters. These hyperparameters can be learned directly from the data, by maximizing the marginal likelihood:
\begin{equation}
    p(y|X,\theta)=\int{p(y|f)p(f|X,\theta) df}
\end{equation}
The likelihood $p(y|f)$ is an immediate consequence of Equation \ref{erm}, and $p(f|X,\theta)$ is the prior distribution defined in Equation \ref{prior}. In the present case, the marginal likelihood is analytically tractable and exact Bayesian inference can be performed \cite{10.5555/1756006.1953029}. In order to make predictions at any location, a new input point $x_*$ is appended to the training data $X$, and the noise variance is incorporated within the upper left block of the covariance matrix. Equation \ref{prior} therefore becomes:
\begin{equation}
\label{joint}
\begin{bmatrix}
y\\
f_*
\end{bmatrix}
\sim\mathcal{N}\Bigg(0,
\begin{bmatrix}
k(X,X)+\sigma^2I_n & k(X,x_*)\\
k(x_*,X) & k(x_*,x_*)
\end{bmatrix}
\Bigg)
\end{equation}
The joint distribution above can then be combined with the marginal likelihood to find the predictive distribution, conditioned on the training data:
\begin{equation}
\label{post}
p(f_*|X,y,x_*)=\frac{p(y,f_*|X,x_*)}{p(y|X)}=\mathcal{N}(f_*|\mu_*(x_*),\Sigma_*(x_*))
\end{equation}
In the present case, the distribution over the predictive output, $f_*$, is Gaussian, and analytical expressions can be found for its mean value $\mu_*$ and variance $\Sigma_*$.
\\\\
More details on GPs and practical implementation considerations can be found in the reference book from Rasmussen \& Williams, \textit{Gaussian Processes for Machine Learning} \cite{10.5555/1756006.1953029}

\subsubsection{Acquisition Functions}

\noindent Gaussian processes are a powerful tool for building a surrogate of the objective function that we want to minimize. As a Bayesian method, it typically avoids the caveat of overfitting \cite{10.5555/922680}, and the predictive distribution offers insight on how much is truly known about the objective function. To find out where the objective function should be evaluated next, acquisition functions are used. The acquisition function embodies a principled means to quantify, for any $x\in\Omega$, how likely it is that evaluating $f(x)$ will yield improvement towards the global minimum of our objective function. Many types of acquisition functions can be found in the literature, but in the current work we adopt \textit{Expected Improvement (EI)}: a popular method that is both analytically tractable and can easily handle constrained BO \cite{frazier2018tutorial, 10.5555/3044805.3044997}.
\\\\
A Bayesian optimization iteration loop must be initialized using a few evaluations of the objective function (typically at random input locations). Among all of the evaluations, the (currently known) minimum is denoted as $\Tilde{f}$. For any $x\in\Omega$, the potential improvement toward the global minimum, given $\Tilde{f}$ is:

\begin{equation}
    I(x)=\max(0, \Tilde{f}-f(x))
\end{equation}

where \noindent $f$ is unknown, but approximated with a GP, so the expected improvement can be derived and expressed in term of the GP predictive mean and variance \cite{frazier2018tutorial}.

\begin{equation}
\label{ei}
\begin{aligned}
    \text{EI}(x)&=\text{E}_{p(f_*|X,y,x)}[I(x)]\\[5pt]
    &=\int_{-\infty}^{+\infty}\max(0,\Tilde{f}-f_*)\mathcal{N}(f_*|\mu_*(x),\Sigma_*(x))df_*\\[5pt]
    &=(\Tilde{f}-\mu_*(x))\Phi(\Tilde{f}|\mu_*(x),\Sigma_*(x))+\Sigma_*(x)\mathcal{N}(\Tilde{f}|\mu_*(x),\Sigma_*(x))
\end{aligned}
\end{equation}

\noindent $\Phi$ refers to the univariate normal cumulative distribution function (CDF). The expected improvement can now be maximized using standard optimization methods (\textit{e.g.} BFGS, gradient descent, \textit{etc.}) \cite{kim2019local}, and the objective function can be evaluated at the new input $x_\text{next}=\arg\max \text{EI}(x)$. Finally, the evaluated value $f(x_\text{next})$ can be used to update the GP predictions, and if a better global minimum candidate value is obtained, the value of $\Tilde{f}$, as well. This process can be iterated until convergence (as gauged by a user specified tolerance) or until the budget for evaluating the objective function has been exhausted. Note that each complete BO iteration comprises two optimization sub-iterations: The first one for maximizing the marginal likelihood when fitting the GP to the data, and the second one for maximizing the acquisition function. As a result, each BO iteration is fairly computationally intensive (but also worth it, when the cost of evaluating the objective function is itself significantly more expensive).
\\\\
In order to deal with constrained optimization context, expressed in Equation \ref{optim}, the standard expected improvement acquisition function in Equation \ref{ei} must be modified \cite{10.5555/3044805.3044997}. To account for the fact that some inputs $x\in\Omega$ are simply not feasible with respect to constraint $c_k(x)\leq\lambda_k$, the modification results in:
\begin{equation}
    cI(x)=\Delta(x)I(x)
\end{equation}
\begin{equation}
\Delta(x)=
\begin{cases}
0\,\,\,\,\text{if}\,\,\,\,c_k(x)\geq\lambda_k\\
1\,\,\,\,\text{if}\,\,\,\,c_k(x)\leq\lambda_k
\end{cases}
\end{equation}
Since $c_k$ may be a black box function, evaluated at the same time as the objective function, it may be approximated with a GP as well. As a result, $\Delta(x)$ is a random variable following a Bernoulli distribution of parameter $\rho(x)$ \cite{10.5555/3044805.3044997}. Furthermore:
\begin{equation}
    \rho(x)=p(c_k(x)\leq\lambda_k)=\int_{-\infty}^{\lambda_k}p(c_{k_*}|X,y,x)dc_{k_*}
\end{equation}
Where $p(c_{k_*}|X,y,x)$ is the Gaussian predictive distribution of $c_{k_*}$ in Equation \ref{post} and conveniently $\rho$ is a univariate Gaussian cumulative distribution function \cite{10.5555/3044805.3044997}. Finally, the expected constrained improvement becomes:

\begin{equation}
\begin{aligned}
    \text{cEI}(x)&=\text{E}[\Delta(x)I(x)]\\[5pt]
    &=\text{E}_{\mathcal{B}(1,\rho(x))}[\rho(x)]\text{E}_{p(f_*|X,y,x)}[I(x)]\\[5pt]
    &=\rho(x)\text{EI}(x)
    \end{aligned}
\end{equation}

\subsubsection{Bayesian Optimization Implementation}

\noindent To implement our Bayesian optimization codes, we use $\texttt{botorch}$ \cite{balandat2019botorch}, a dedicated python library. $\texttt{botorch}$ relies on $\texttt{gpytorch}$ \cite{gardner2018gpytorch} for performing all the relevant GP inferences, and both of these libraries are built on top of the deep learning framework $\texttt{pytorch}$ \cite{paszke2017automatic}.

\section{FSI Validation} \label{Turek}
\begin{figure}[h]
\centering
\includegraphics[width=1\textwidth]{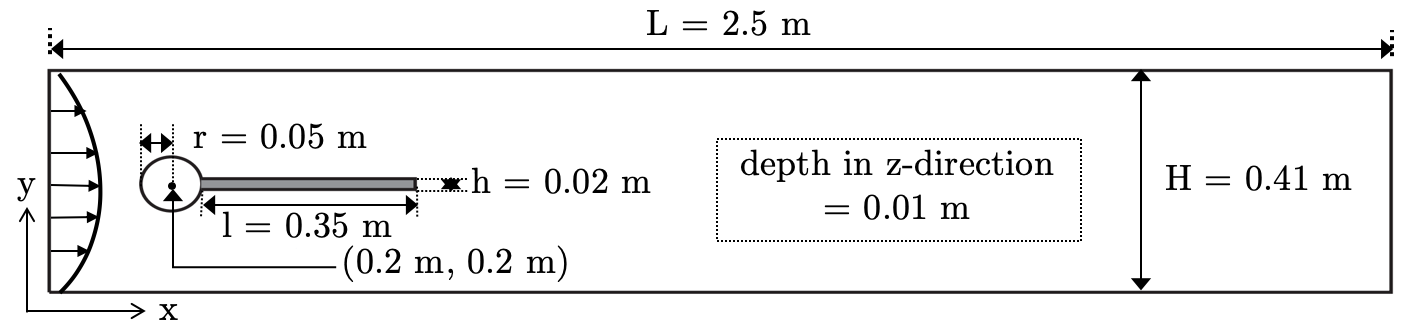}
\caption{Turek and Hron FSI3 case}\label{th_struc}
\end{figure}

\noindent As part of our design approach, we carefully searched the literature for a canonical FSI verification problem that was similar enough to our ultimate design application (\textit{i.e.} fiber reinforced plastic UUAV sail plane), to be useful is building or confidence in the veracity of our FSI modeling approach. To that end, the Turek \& Hron's benchmark cases \cite{turek2006} are considered here as we judge the accuracy of our implicit partitioned FSI approach using CU-BENs and OpenFOAM solvers. The general problem setup for our verification context is shown in Figure \ref{th_struc}. Within the structural domain, an elastic cantilever beam is attached to a fixed cylinder whose radius is 5 cm; the center of the cylinder is positioned at (0.2 m, 0.2 m). The cantilever beam is 0.35~m long, 0.02~m thick, and 0.01~m wide. \\%
\newline
No-slip boundary conditions are imposed at the top and bottom walls of the channel, as well as the interface along the cylinder and the cantilever. The pressure outflow condition along the right edge of the channel is prescribed to atmospheric pressure. The channel inflow condition is defined along the left edge of the domain as a parabolic velocity profile, described in Equation \ref{th_eq1}, where U is the mean inflow velocity. Equation \ref{th_eq2} describes the evolution of the inflow velocity as a function of time. \\\begin{equation} \label{th_eq1}
u_f (0, y) = 1.5U\frac{4}{0.1681}y(0.41-y)
\end{equation}
\begin{equation} \label{th_eq2}
u_f(t,0,y)=\begin{cases}
\displaystyle\frac{u_f(0,y)[1-\cos(\frac{\pi t}{2})]}{2} & \text{if $t<2.0$}\\
u_f(0,y)  & \text{otherwise}
\end{cases}
\end{equation}
The verification procedure sets out to verify three test cases (\emph{i.e.} the CSM3 test, the CFD3 test, and the FSI3 test) detailed in \cite{turek2006}, each test case evaluates the behavior of the corresponding component response within our CU-BENs/OpenFOAM implicit FSI solver, so that comparisons can be made with the ground truth data reported in \cite{turek2006}. Mesh refinement studies are undertaken at this point, so that we may know what spatiotemporal descretizations are suitable as we move towards our ultimate design case.
\subsection{CSM3 Test}
\begin{figure}[h]
\centering
\includegraphics[width=1\textwidth]{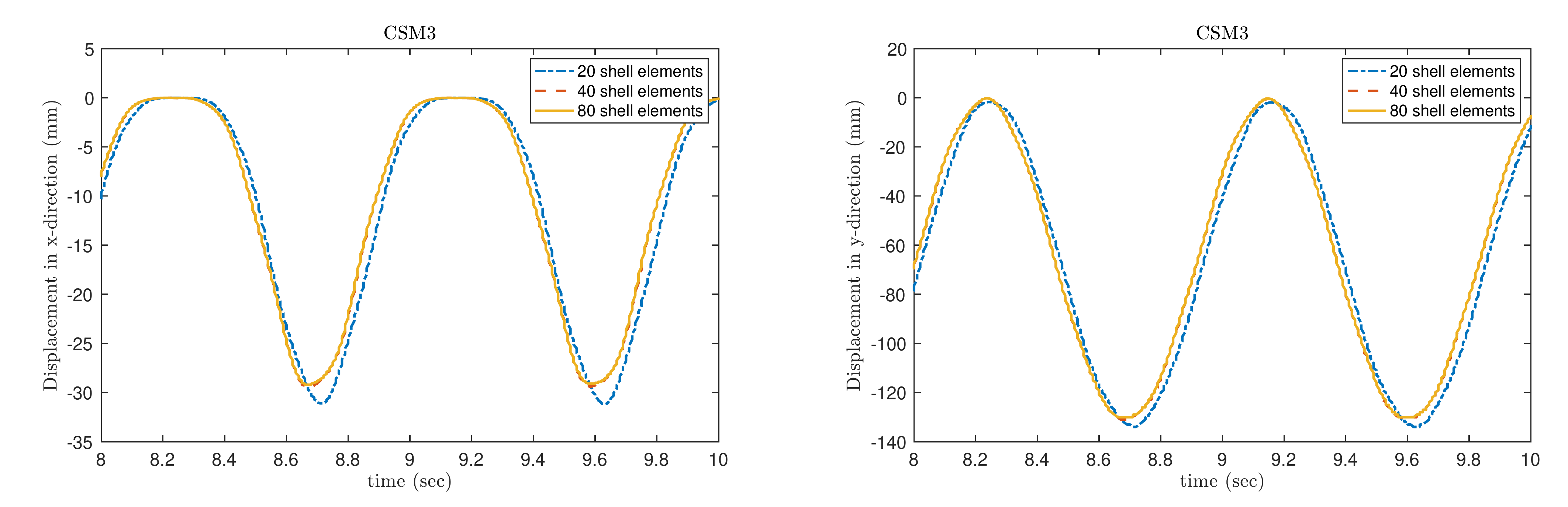} \\
\caption{x-- and y-- displacements at the free end of the cantilever at time $t = [8, 10] $ seconds: $\delta t= 0.005~\text{sec}$} \label{th_csd}
\end{figure}

\noindent An elastic cantilever beam subjected to gravitational force, only, with $g = 2~\text{m/s}^2$ is carried out in the CSM3 test. The elastic beam has a Young's modulus of 1.4~MPa, Poisson ratio of 0.4, and mass density of $1000~\text{kg/m}^3$. Implicit Newmark time integration scheme, coupled with Newton Raphson sub-iteration within each time step, is used to capture the geometric nonlinearity effects within the CU-BENs structural model. The displacements at the free end of the cantilever are recorded. Figure \ref{th_csd} and Table \ref{csd_disp} presents the x-- and y-- tip displacements over the time span $t = [8, 10] $ seconds. As shown, that the tip displacements tend to the reference values provided by Turek \& Hron, as the number of shell elements increases; negligible differences among the displacements are observed. \\
\begin{table}[h]
\centering
\resizebox{\textwidth}{!}{\begin{tabular}{c|c|c|c|c}
Number of shells & x-- displacement [centerline  $\pm$ amplitude] & frequency [x-- disp]& y-- displacement [centerline  $\pm$ amplitude]  & frequency [y-- disp]\\ \hline
20 & -15.602  $\pm$  15.598 & 1.087 &-67.940 $\pm$  66.060 & 1.087   \\  \hline
40 &  -15.002 $\pm$ 14.998 & 1.0989 &-66.441 $\pm$ 65.560 & 1.0989\\ \hline
80 &  -14.552 $\pm$ 14.547 & 1.0989 &-65.152 $\pm$ 64.848 & 1.0989\\ \hline
Turek \& Hron reference & -14.305 $\pm$  14.305 &1.0995 &-63.607 $\pm$  65.160 & 1.0995 \\
\end{tabular}}
\caption{Peak x-- and y-- displacement at the free end of the cantilever at time $t = [8, 10] $ seconds: $\delta t= 0.005~\text{sec}$}
\label{csd_disp}
\end{table}
\subsection{CFD3 Test}
\begin{figure}[h]
\centering
\includegraphics[width=1\textwidth]{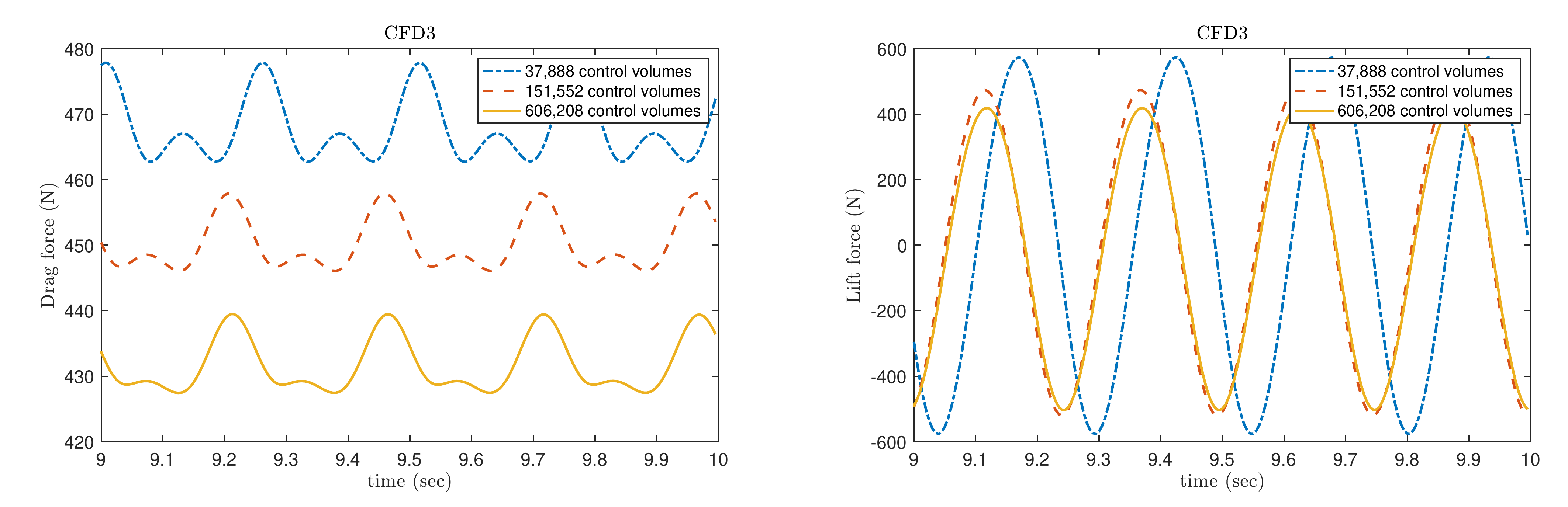} \\
\caption{Drag and lift forces along the cylinder and the cantilever at time $t = [9, 10] $ seconds: $\delta t= 0.005$~sec}\label{th_cfd}
\end{figure}

\noindent Next, the fluid dynamic aspects of the coupled system are verified using the CFD3 test case, within which the cantilever is modeled as a rigid beam. A parabolic inflow described in Equation \ref{th_eq2}, with 2~m/s mean inflow velocity, is prescribed along the left boundary edge of the channel. The fluid flow is simulated with OpenFOAM's incompressible solver: PimpleDyMFoam. The total drag and lift forces along the cylinder, as well as along the rigid beam, are recorded. The drag and lift over the time span $t = [9, 10] $ seconds can be found in Figure \ref{th_cfd} and Table \ref{cfd_force}. We see that the drag and lift forces computed using PimpleDyMFoam are converging to Turek \& Hron's values with increasing number of control volumes. At the finest level of refinement, with 606,208 control volumes, the percentage differences for the peak drag and lift values are well below 2 \%. \\
\begin{table}[h]
\centering
\resizebox{\textwidth}{!}{\begin{tabular}{c|c|c|c|c}
Number of control volumes & drag force [centerline $\pm$ amplitude] &  frequency [drag]& lift force [centerline $\pm$ amplitude]  & frequency [lift]\\ \hline
37,888  & 470.29  $\pm$ 6.915 & 3.42 & -1.2882 $\pm$  574.86 &  3.9216\\  \hline
151,552 & 451.98  $\pm$  8.285 &  4  &-22.998 $\pm$  496.76 &  4 \\  \hline
606,208 &  431.90 $\pm$ 4.3290 & 4 & -42.279  $\pm$ 461.54 &  4 \\ \hline
Turek \& Hron reference & 439.45 $\pm$  5.6183 &4.3956 &-11.893 $\pm$  437.81 & 4.3956 \\
\end{tabular}}
\caption{Peak drag and lift forces along the cylinder and the cantilever at time $t = [9, 10] $ seconds: $\delta t= 0.005$~sec}
\label{cfd_force}
\end{table}
\subsection{FSI3 Test}
\begin{figure}[h]
\centering
\includegraphics[width=1\textwidth]{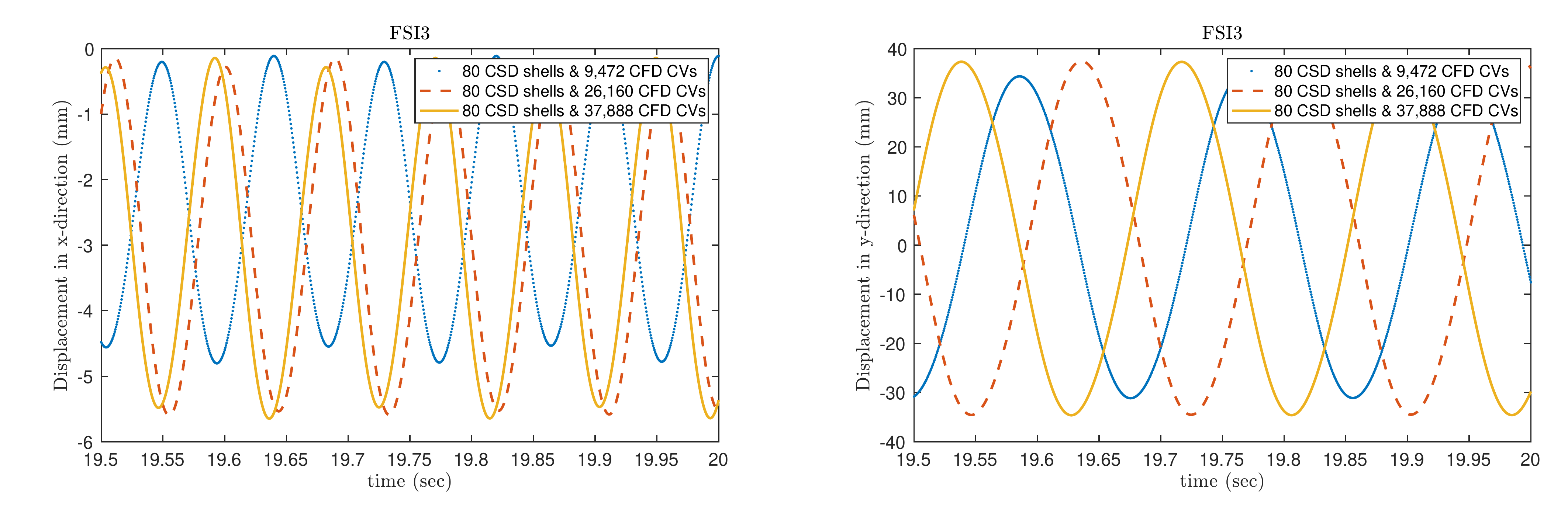} \\
\caption{x-- and y-- displacements at the free end of the cantilever during $t = [19.5, 20] $ seconds: $\delta t= 0.0005$~sec} \label{fsi3_disp1}
\end{figure}

\begin{figure}[h]
\centering
\includegraphics[width=1\textwidth]{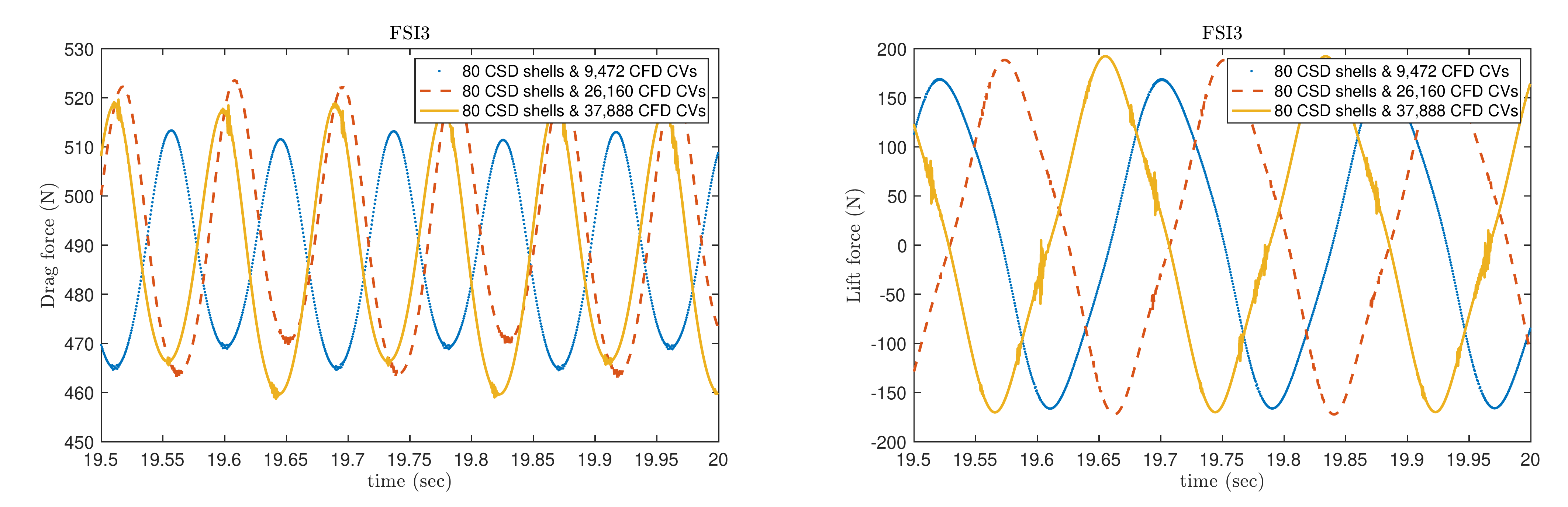} \\
\caption{Drag and lift forces along the the cantilever during time $t = [19.5, 20] $ seconds: $\delta t= 0.0005$~sec} \label{fsi3_force1}
\end{figure}

\noindent The complete FSI modeling context is treated in FSI3 test case, where an elastic structure (Young's modulus of 5.6~MPa, Poisson ratio of 0.4, and mass density of 1000~kg/m$^3$) is submerged in a fluid channel that is 2.5~m long, 0.41~m tall, and 0.01~m wide; a summary of the material properties for the fluid and structural domains is provided in Table \ref{th_FSItb}. The channel inlet, on the left, is prescribed with a parabolic inflow speed with mean inflow velocity of 2~m/s. The fluid flow is modeled with OpenFOAM's PimpleDyMFoam solver, while, the structural deformation is estimated using the Generalized-$\alpha$ integration method within CU-BENs (a spectral radius of 0.8 is applied to ensure numerical stability.) Deformation along the fluid and structural interface is governed by the IQN-ILS scheme. Displacements along the moving boundary are iterated until the residual vector ($\mathbf{R}^{k}$) satisfies the convergence criterion $\left\|\mathbf{R}^{k}\right\|\leq\epsilon$, where the convergence tolerance is $\epsilon = 10^{-5}$.  \\
\begin{table}[h]
\caption{Engineering properties}\label{th_FSItb}
\centering
\begin{tabular}{ l l l l } \\
\hline
 Fluid Properties & & Structural Properties & \\
   \hline
  Density, $\rho_f$ & $10^3~\text{kg/m}^3$ & Density, $\rho_s$ & $10^3~\text{kg/m}^3$ \\
  Viscosity, $\mu_f $ &   $10^{-3}~\text{m}^2\text{/s}$ & Poisson ratio, $\mu_s$ & 0.4\\
  Reynolds number, Re & 200 & Young`s modulus, E & 5.6 MPa \\
  \hline
\end{tabular}
\end{table} \\
The full analysis time (not wall clock) is 20 sec with $\delta t= 0.0005$~sec. The x-- and y-- tip displacements over the time span $t = [19.5, 20] $ seconds as well as the peak x-- and y-- tip displacements are provided in Figure \ref{fsi3_disp1} and Table \ref{fsi3_disp2}. The analysis is repeated using three levels of mesh refinement in the fluid domain, meanwhile the discretization in the structural domain is fixed at 80 DKT shell elements. As shown, the implicit FSI solver is able to predict the structural deformation reasonably well compared to the reference values provided by Turek \& Hron. The peak displacements are slightly higher than the benchmark values with the peak x-- displacement converging towards $ -2.89 \pm 2.75$~mm and the peak y-- displacement converging towards $1.36 \pm 35.97$~mm. Nonetheless, the discrepancies among the peak displacements are within $1$~mm in x-- direction and $2$~mm in y-- direction, approximately $12.5 \%$ difference for x-- displacement and $4.1 \%$ difference for y-- displacement.\\
\begin{table}[h]
\centering
\resizebox{\textwidth}{!}{\begin{tabular}{c|c|c|c|c|c}
Number of shells & Number of control volumes & x-- displacement [centerline $\pm$ amplitude] & frequency [x-- disp]& y-- displacement [centerline  $\pm$ amplitude]  & frequency [y-- disp]\\
80 & 9,472 & -2.46  $\pm$  2.35 & 11.11 & 1.61 $\pm$  32.75 & 5.56   \\  \hline
80 &  26,160 & -2.87 $\pm$ 2.73 & 11.11 & 1.46 $\pm$ 36.0 & 5.56 \\ \hline
80 &  37,888 & -2.89 $\pm$ 2.75 & 11.11 & 1.36 $\pm$ 35.97 & 5.56 \\ \hline
\multicolumn{2}{c}{ Turek \& Hron reference} &  -2.69 $\pm$  2.53 &10.9 & 1.48 $\pm$  34.38 & 5.3 \\
\end{tabular}}
\caption{Peak x-- and y-- displacements at the free end of the cantilever during time $t = [19.5, 20] $ seconds: $\delta t= 0.0005$~sec}
\label{fsi3_disp2}
\end{table} \\
The drag and lift responses over the time span  $t = [19.5, 20]$ seconds are presented in Figure \ref{fsi3_force1} and Table \ref{fsi3_force2}. Minor nonphysical oscillations can be observed due to the different spatial discretization in the fluid and structural systems as well as nonconforming mesh along the fluid-structure interface. The implicit FSI solver estimates the peak drag converge towards $489.2  \pm 30.44$~N,  and the peak lift converge towards $11.084  \pm 181.21$~N. The discrepancies among the peak drag and lift are within $40$~N and $170$~N respectively, with percentage difference of $8 \%$ for drag and $26.5 \%$ for lift. The discrepancies are due to the difference in fluid solver used in the present work (\emph{i.e.} different spatial temporal discretization scheme, different approach to solving the pressure-velocity equations, \textit{etc.}) as well as the different FSI approach compared to Turek \& Hron. As specifically noted in \cite{turek2010}, with different discretization, solver, and coupling mechanism, differences in different approach could lead to discrepancies of up to  $50 \%$ for the drag and lift values and $10 \%$ difference for the displacement values. With this in mind, we are confident in our FSI predictions, as we are well below these thresholds.\\

\begin{table}[h]
\centering
\resizebox{\textwidth}{!}{\begin{tabular}{c|c|c|c|c|c}
Number of shells & Number of control volumes & drag force [centerline  $\pm$ amplitude] &  frequency [drag]& lift force [centerline $\pm$ amplitude]  & frequency [lift]\\
80 & 9,472  & 488.9 $\pm$ 24.40 & 11.11 & 1.399 $\pm$  167.70 &  5.56 \\  \hline
80 & 26,160 & 493.3  $\pm$  30.28 &  11.11 & 8.104 $\pm$  180.34 &  5.56 \\  \hline
80  & 37,888 &  489.2  $\pm$ 30.44 & 11.11  & 11.084  $\pm$ 181.21&  5.56 \\ \hline
\multicolumn{2}{c}{ Turek \& Hron reference} & 457.3 $\pm$  22.66 & 10.9 & 2.22 $\pm$  149.78 & 5.3  \\
\end{tabular}}
\caption{Peak drag and lift forces during time $t = [19.5, 20] $ seconds: $\delta t= 0.0005$~sec}
\label{fsi3_force2}
\end{table}

\section{Bridging Simulations}\label{examples}
\begin{figure}[ht!]
\centering
\includegraphics[width=1\textwidth]{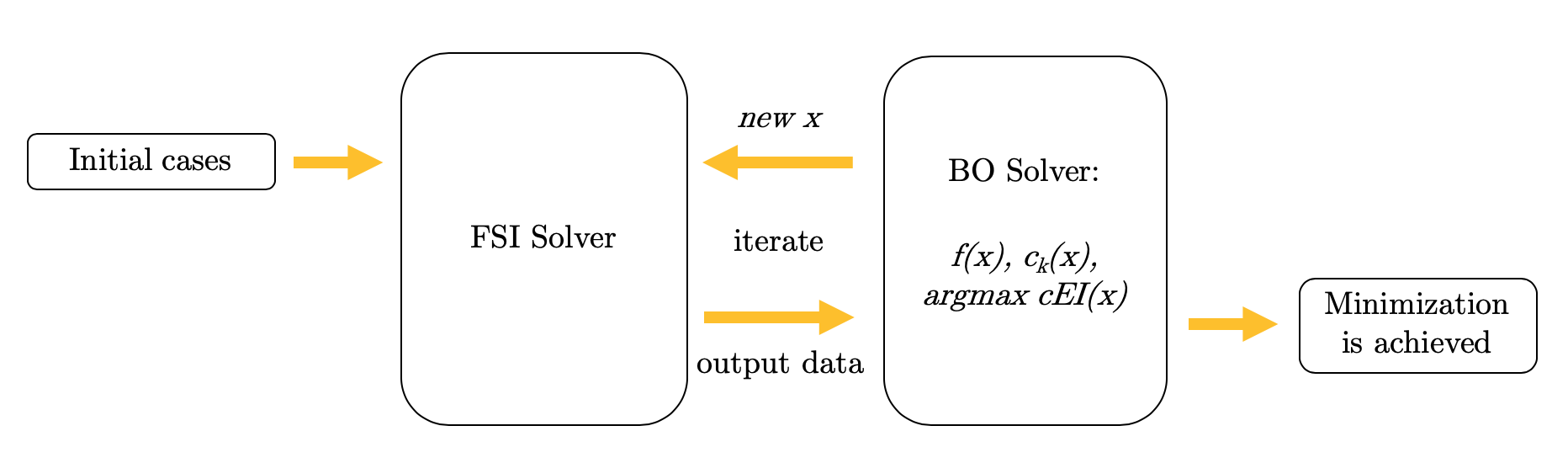} \\
\caption{A schematic representation of the interplay between the FSI and Bayesian optimization solvers} \label{bo_fsi}
\end{figure}

\noindent In the previous section, we identified canonical verification models to furnish ground truth results for assessing the performance of our proposed FSI simulation framework. These versification analyses allowed us to identify suitable spatiotemporal discretization for our application space, along with gauging the efficacy of our selected numerical schemes. As it is that we need to extend our analyses to be more realistic with respect to our ultimate design application (fiber reinforced polymer UUAV sail plane), we now incrementally increase our modeling complexity using what refer to as \textit{bridging simulations}. We perform Bayesian optimization (both constrained and unconstrained) using our bridging simulations. The geometric configuration and boundary conditions of the FSI3 benchmark, outlined in Section \ref{Turek}, are leveraged in the bridging simulations, as detailed in Section \ref{example1} to \ref{example3}. As it is that our ultimate design application involves the material property design for a composite fiber reinforced structural component of fixed geometry, in our bridging simulations the elastic modulus of our structure is specified as a function of its cantilevered geometry within our bridging simulations: as linear, uniform, and box functions, respectively. A schematic representation of the BO-FSI procedure is presented in Figure \ref{bo_fsi}, where $x$ is the optimization parameter (also referred to as input), $f(x)$ is the objective function, $c_k(x)$ is the set of constraint functions, and $cEI(x)$ is the acquisition function. The BO process goes as follows:
\begin{enumerate}
    \item We initialize the optimizer by defining a set of initialization inputs (chosen randomly except in section \ref{example3} and \ref{UUAV}).
    \item We pass the initialization inputs to the FSI solver, which outputs corresponding data on the objective function and the constraints.
    \item The BO solver computes the GP surrogates, maximizes the acquisition function, and provide a new input point.
    \item The new input point is passed to the FSI solver, and the process it repeated until the desired design is uncovered.
\end{enumerate}

\noindent Note that in Bayesian optimization, there is no obvious, \textit{a priori} stopping criterion; though there may be a hard limit on the number of feasible iterations (\textit{e.g.} available computational budget for evaluating the objective function): knowing when to stop may be difficult. There is a trade-off to balance between objective function evaluation cost and the likelihood of finding a better (the global) minimum. The optimizer is typically stopped when no significant progress has been made for the last few iterations, or if the uncovered design is simply ``good enough'' for engineering purposes.

\subsection{Example 1: Minimizing Tip Displacement with Outlet Velocity constraint} \label{example1}

\noindent In this first example, we aim to minimize the vertical tip displacement of the cantilever beam, used in the FSI3 verification, subject to an outlet velocity constraint. In this section, and those subsequent, the tip displacement corresponds to the maximum displacement attained at any time point within the stabilized FSI run. The beam elasticity modulus, $E$, is assumed to  vary linearly along the length of the cantilever (with zero coordinate corresponding to the fixed end, at the cylinder):
\begin{equation}
\begin{cases}
    E(x)=Ax+B\,\,\,\,\,x\in[0,l]\\[5pt]
    \displaystyle\int_0^lE(x)dx=1.96~\text{MPa}
\end{cases}
\label{ex1_E}
\end{equation}
The integral of the elastic modulus function, taken along the beam, remains constant for all time; this results in $B$ being dependant on $A$. In addition, to ensure $E(x)>0$, $A$ is bounded in the interval $[-31.9088, 31.9088]$. Our goal is to determine the value of $A$, such that the vertical beam tip displacement, $\delta$, is minimized, while also making sure that we do not exceed our prescribed maximum outlet fluid velocity, $v_c=1.4\text{\,m/s}$, where the velocity is taken at mid-height. This example scenario typically relates to situations where making the beam stiffer is costly, and thus choices must be made on where to concentrate the stiffness throughout the beam in an optimal manner. Formally, the optimization problem is:

\begin{equation}
    \text{Find}\,\,\,\,\,\min_A\delta\,\,\,\,\,\text{s.t.}\,\,\,\,\,v_x\leq v_c
\end{equation}

\begin{figure}[ht!]
\centering
\captionsetup[subfigure]{justification=centering}
\subfloat[Initialization\label{ex1iter1}]{\includegraphics[width=0.33\textwidth]{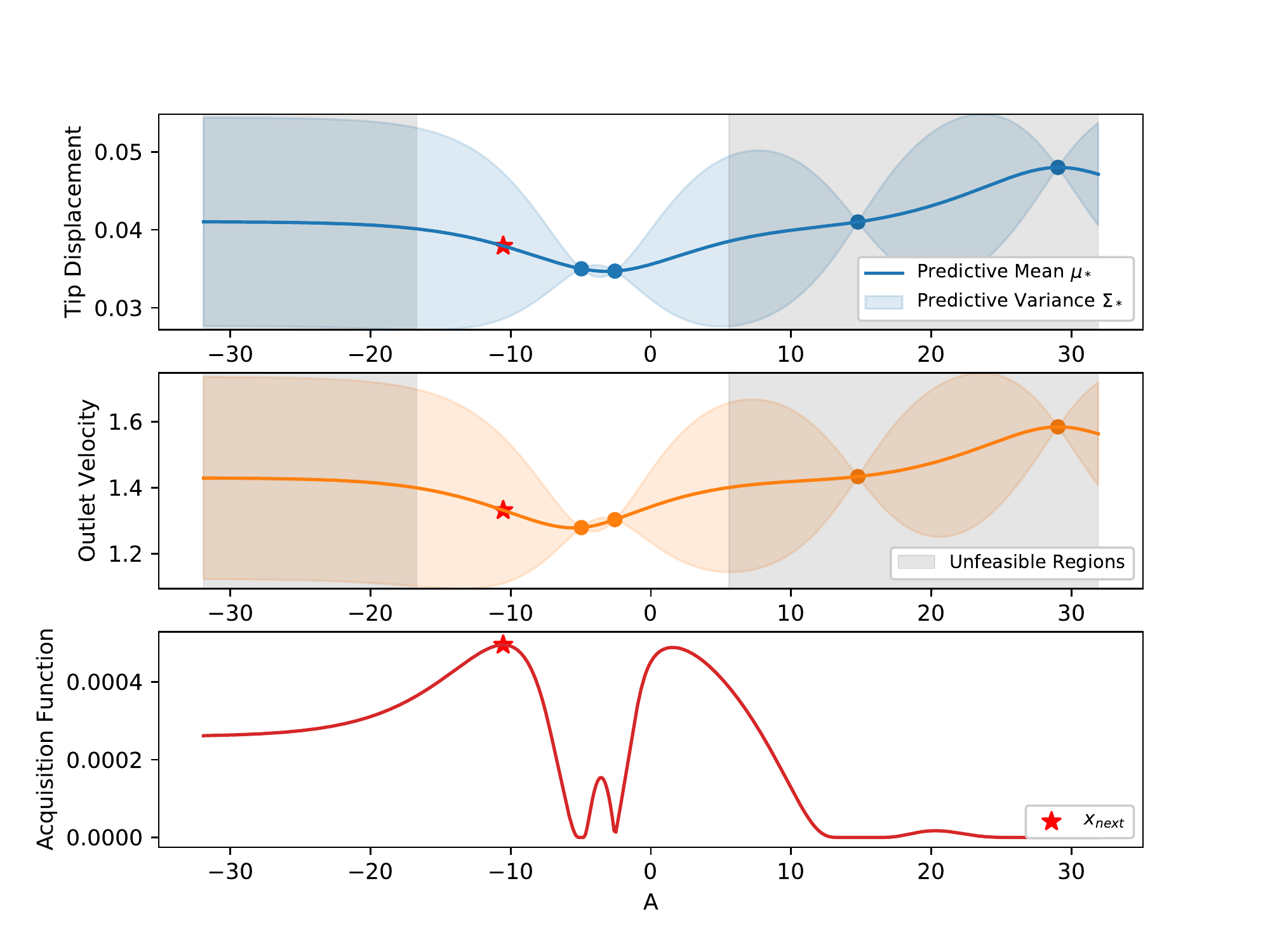}}\hfill
\subfloat[Iteration 1\label{ex1iter2}] {\includegraphics[width=0.33\textwidth]{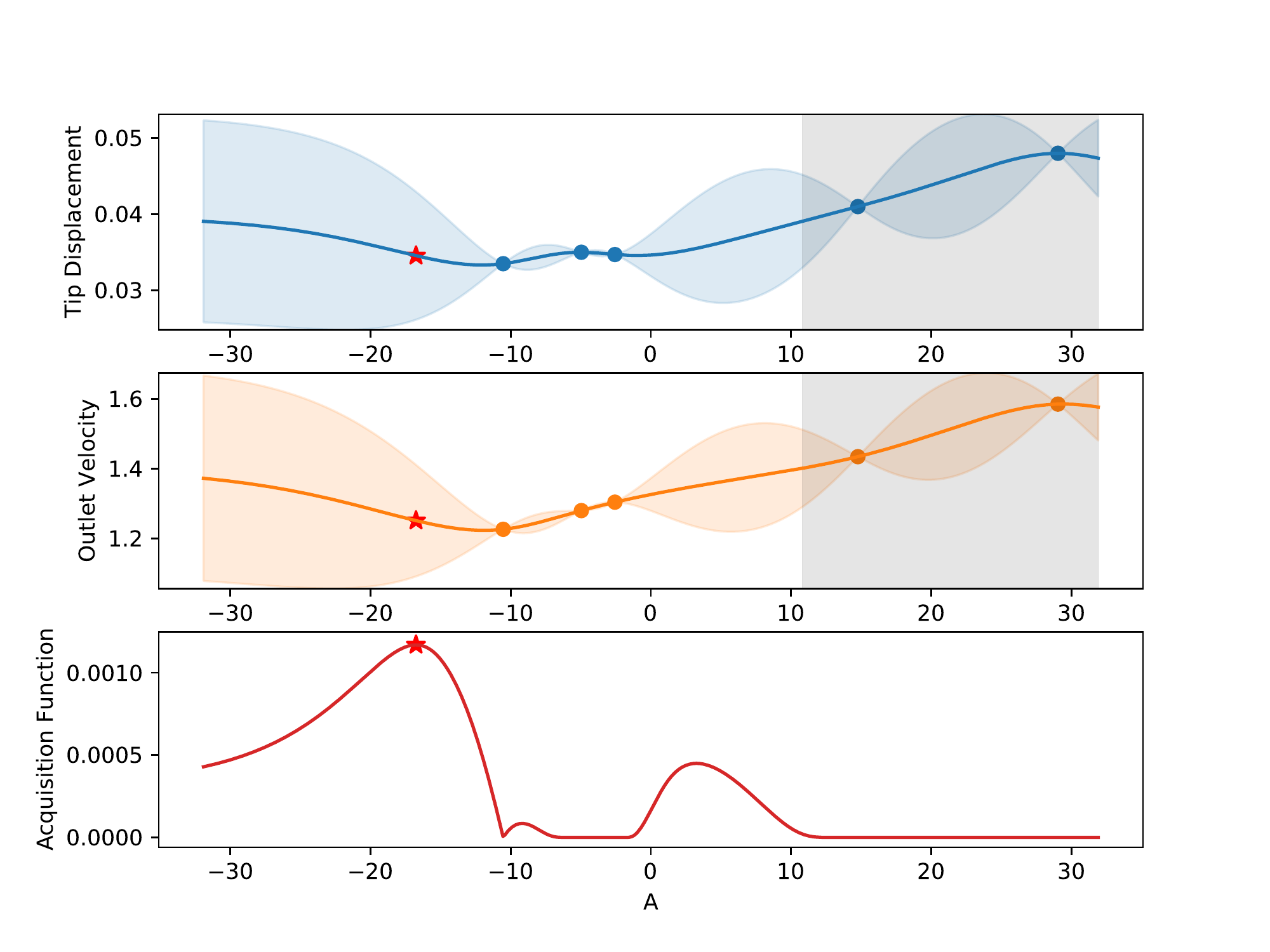}}\hfill
\subfloat[Iteration 2\label{ex1iter3}]{\includegraphics[width=0.33\textwidth]{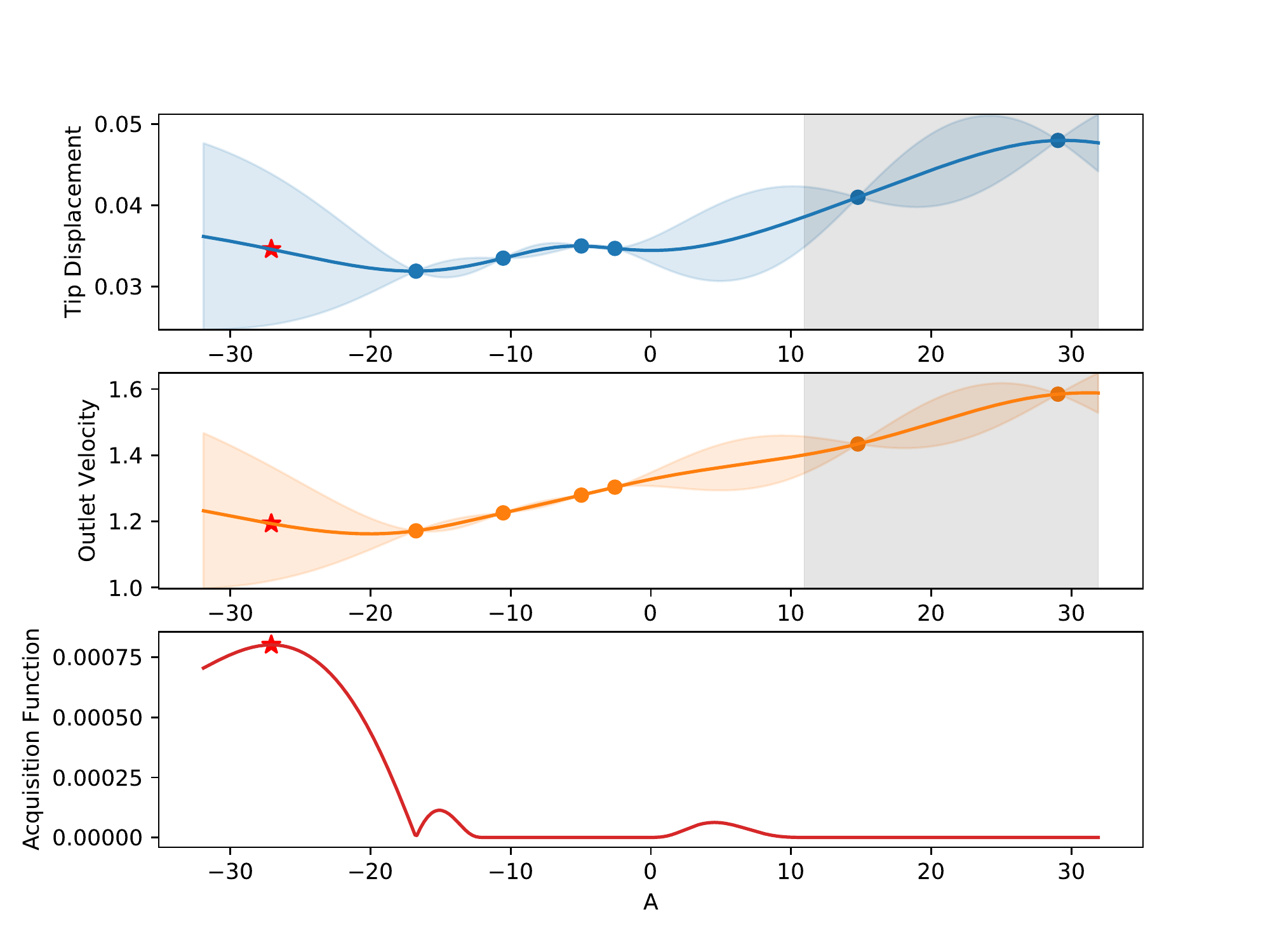}}\hfill
\subfloat[Iteration 3\label{ex1iter4}]{\includegraphics[width=0.33\textwidth]{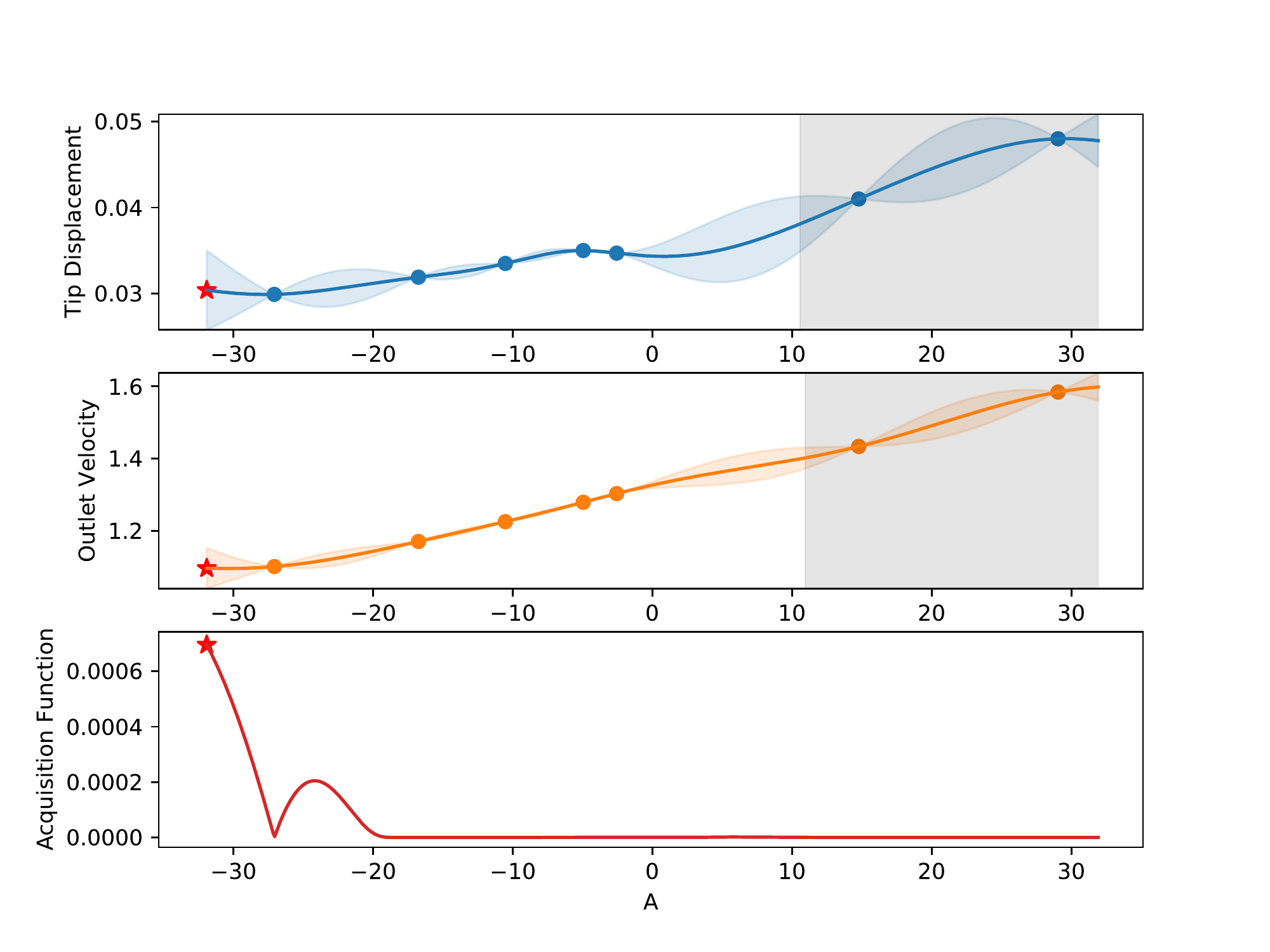}}\hfill
\subfloat[Iteration 4\label{ex1iter5}]{\includegraphics[width=0.33\textwidth]{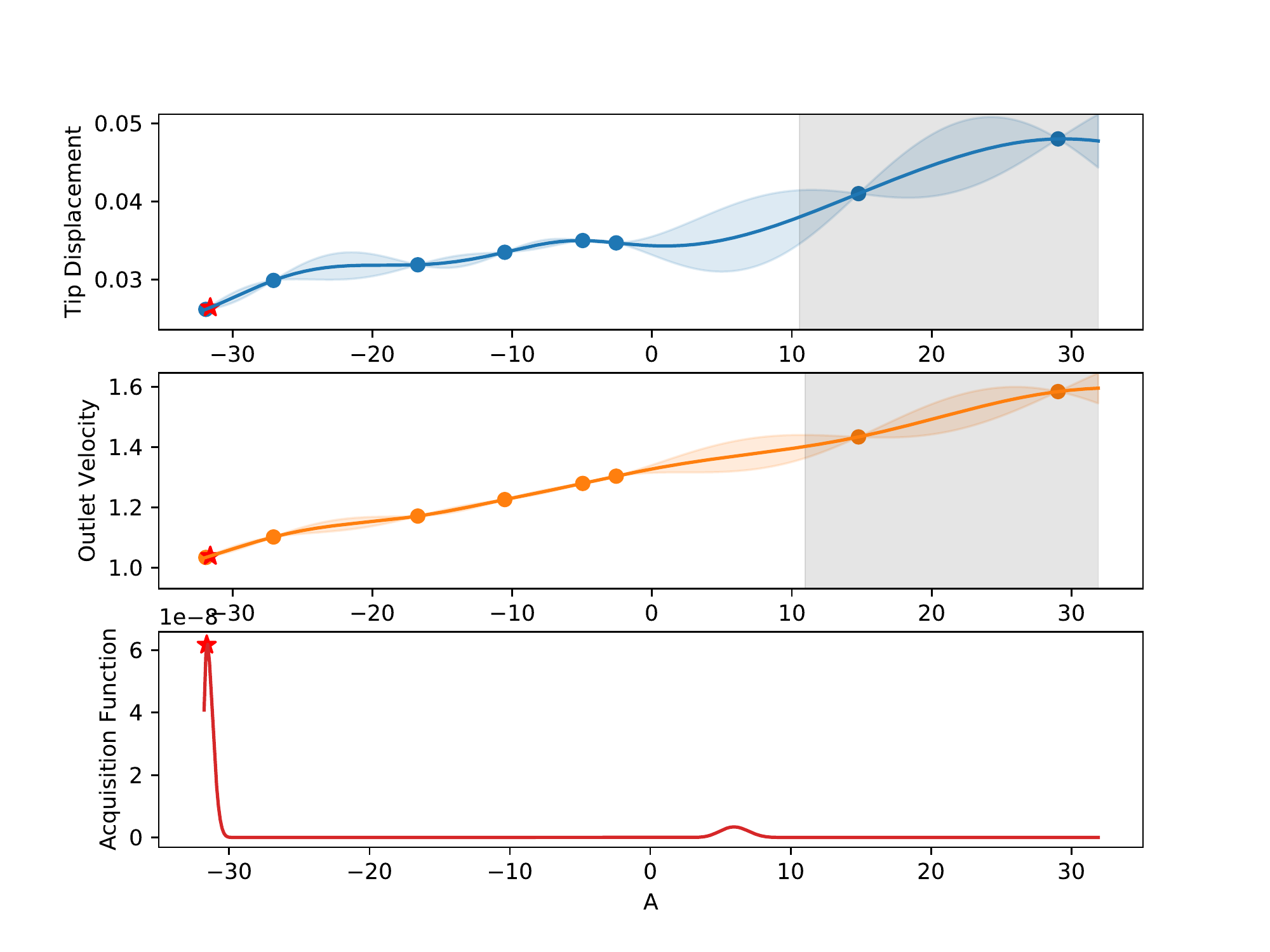}}\hfill
\subfloat[Iteration 5\label{ex1iter6}]{\includegraphics[width=0.33\textwidth]{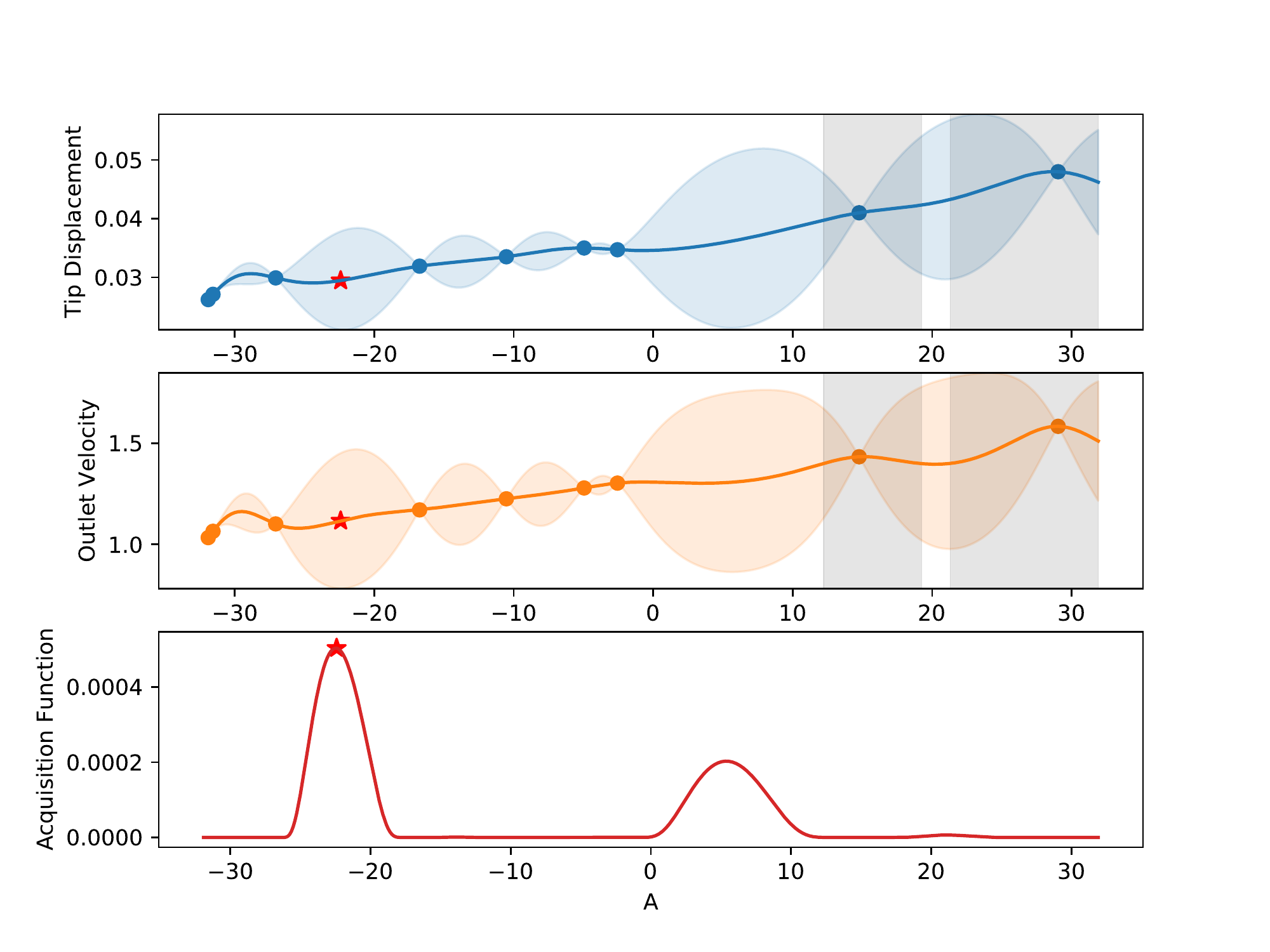}}
\caption{Successive BO iterations - tip displacement with outlet velocity constraint. The solid blue and orange lines represent the predictive mean of the two surrogate GPs ($\delta$ and $v_x$ respectively) with their $95\%$ confidence interval (shaded areas), given the already known tip displacements and outlet velocity values. The acquisition function indicates what value of $A$ is the most likely to yield a better global minimum (\textit{i.e.} the value that should be use in the next FSI run).} \label{ex1iter}
\end{figure}

\begin{figure}[h!]
    \centering
    \includegraphics[width=0.5\textwidth]{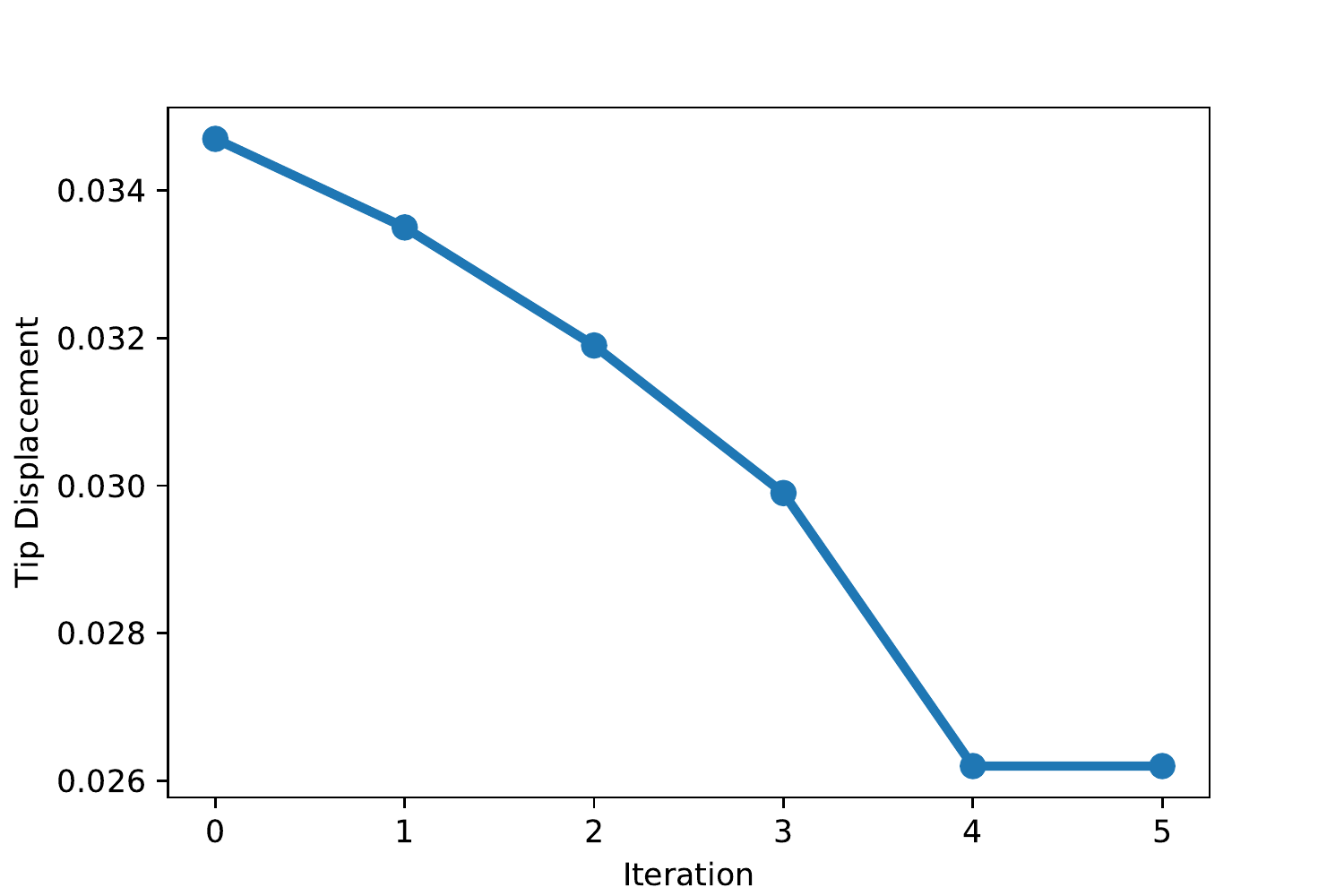}
    \caption{Best minimum recorded at each BO iteration - tip displacement with outlet velocity constraint}
    \label{ex1descent}
\end{figure}

\noindent The Bayesian optimizer is initialized using four different FSI runs; instantiated with random values for A. Table \ref{ex1values} shows the successive input values proposed by the optimizer, as well as the corresponding tip displacement and outlet velocity. Figure \ref{ex1iter} shows plots of the surrogate GPs for $\delta(A)$ and $v_x(A)$, along with the constrained expected improvement at the initialization, and after each BO iteration. Ultimately, the tip displacement, and the outlet velocity, both appear to have a monotonic trend; thus the optimizer pushes us towards the lower bound of $A$ (\textit{i.e.} where the stiffness is concentrated at the beam's fixed end). Once $(A_\text{min}=-31.9088, \delta_\text{min}=0.0262)$ has been reached, no progress towards a better global minimum is being made, and given the likely monotonic trend of the objective function, future progress is deemed unlikely, and so the optimizer is stopped (Figure \ref{ex1descent}).

\begin{table}[h]
\centering
\begin{tabular}{c|c|c|c}
Iteration \# & $A$ (MPa)& $\delta$ (m)& $v_x$ (m/s)\\\hline
Init & $-2.565$ & $0.0347$ & $1.3035$\\\hline
Init & $-4.957$ & $0.035$  & $1.2794$\\\hline
Init & $29.033$ & $0.048$ & $1.5844$\\\hline
Init & $14.764$ & $0.0410$ & $1.4338$\\\hline
1 & $-10.5329$ & $0.0335$ & $1.2258$ \\\hline
2 & $-16.7519$ & $0.0319$ & $1.171$\\\hline
3 & $-27.0711$ & $0.0299$ & $1.1016$\\\hline
\textbf{4} & $\mathbf{-31.9088}$ & $\mathbf{0.0262}$ & $\mathbf{1.0337}$\\\hline
5 & $-31.5656$ & $0.0271$ & $1.0651$\\
\end{tabular}
\caption{\label{ex1values} Successive BO iterations - tip displacement with outlet velocity constraint}
\end{table}

\subsection{Example 2: Unconstrained Uniform Stiffness} \label{example2}

\noindent In this second example, we now aim to minimize the vertical tip displacement of the cantilever with respect to $B$ in Equation \ref{ex1_E}, assuming that the beam has an uniform stiffness ($A=0$) (\emph{i.e.} integral condition on $E(x)$ is no longer enforced). $B$ is bounded in the interval $[0, 10]$, but no additional constraints are used. This leads to an unconstrained BO, with the standard expected improvement acquisition function:
\begin{equation}
    \text{Find}\,\,\,\,\,\min_B\delta
\end{equation}

\begin{figure}[ht]
\centering
\captionsetup[subfigure]{justification=centering}
\subfloat[simulation time = $6.25$~sec; \newline B= 5.6]{\scalebox{0.8}{\includegraphics[width=0.5\textwidth]{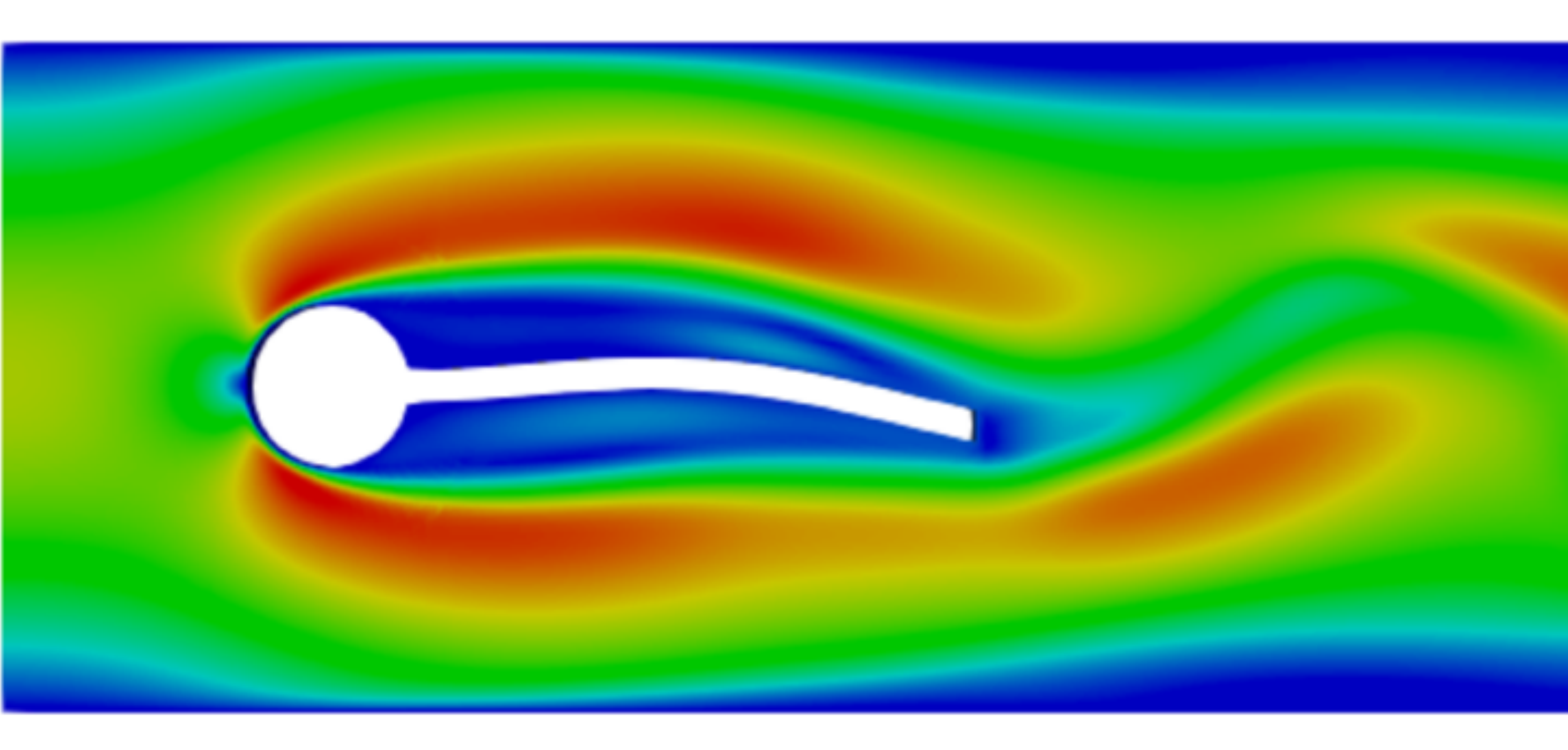}}}
\subfloat[simulation time = $6.25$~sec; \newline B= 2]{\scalebox{0.8}{\includegraphics[width=0.5\textwidth]{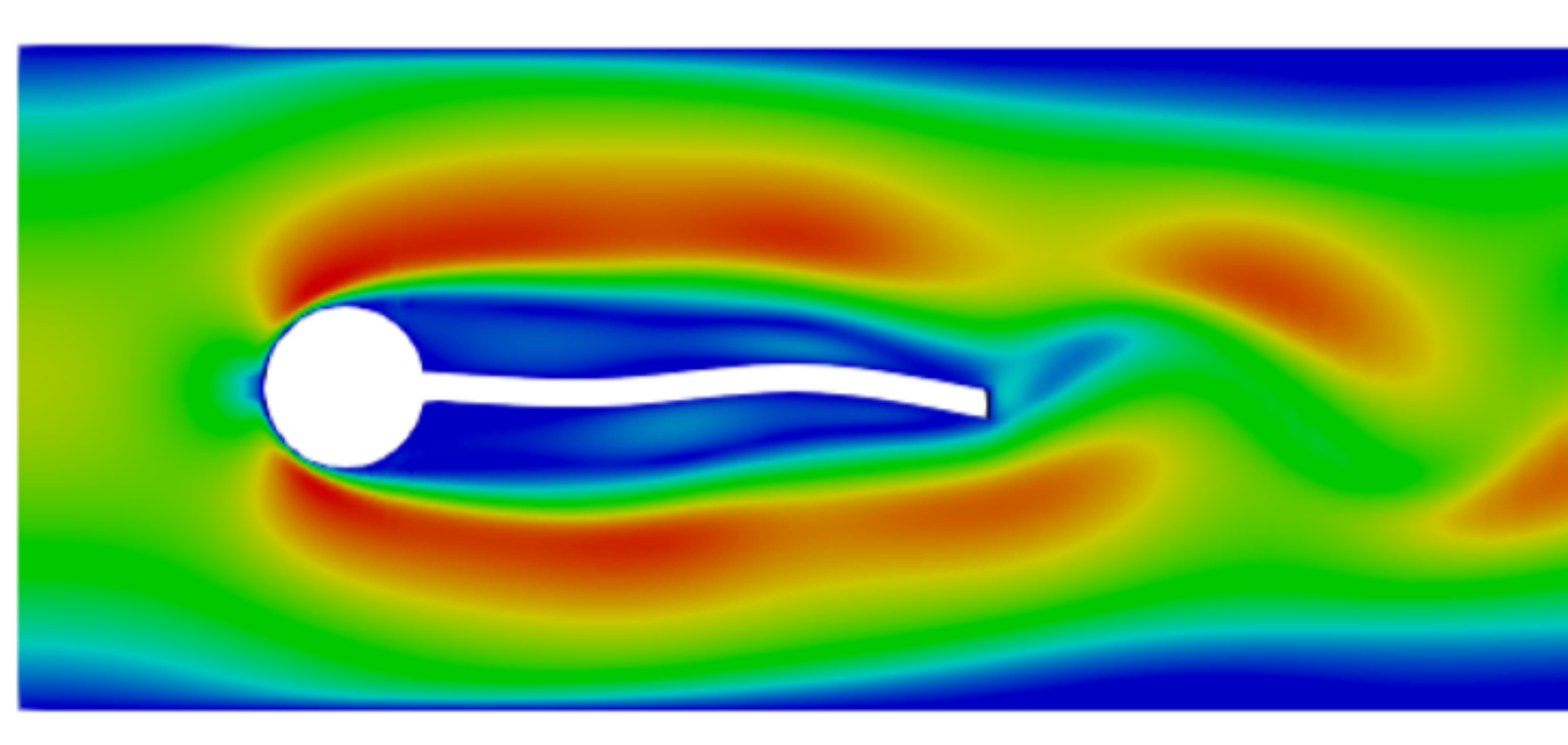}}}\\
\subfloat[simulation time = $12.5$~sec; \newline B= 5.6]{\scalebox{0.8}{\includegraphics[width=0.5\textwidth]{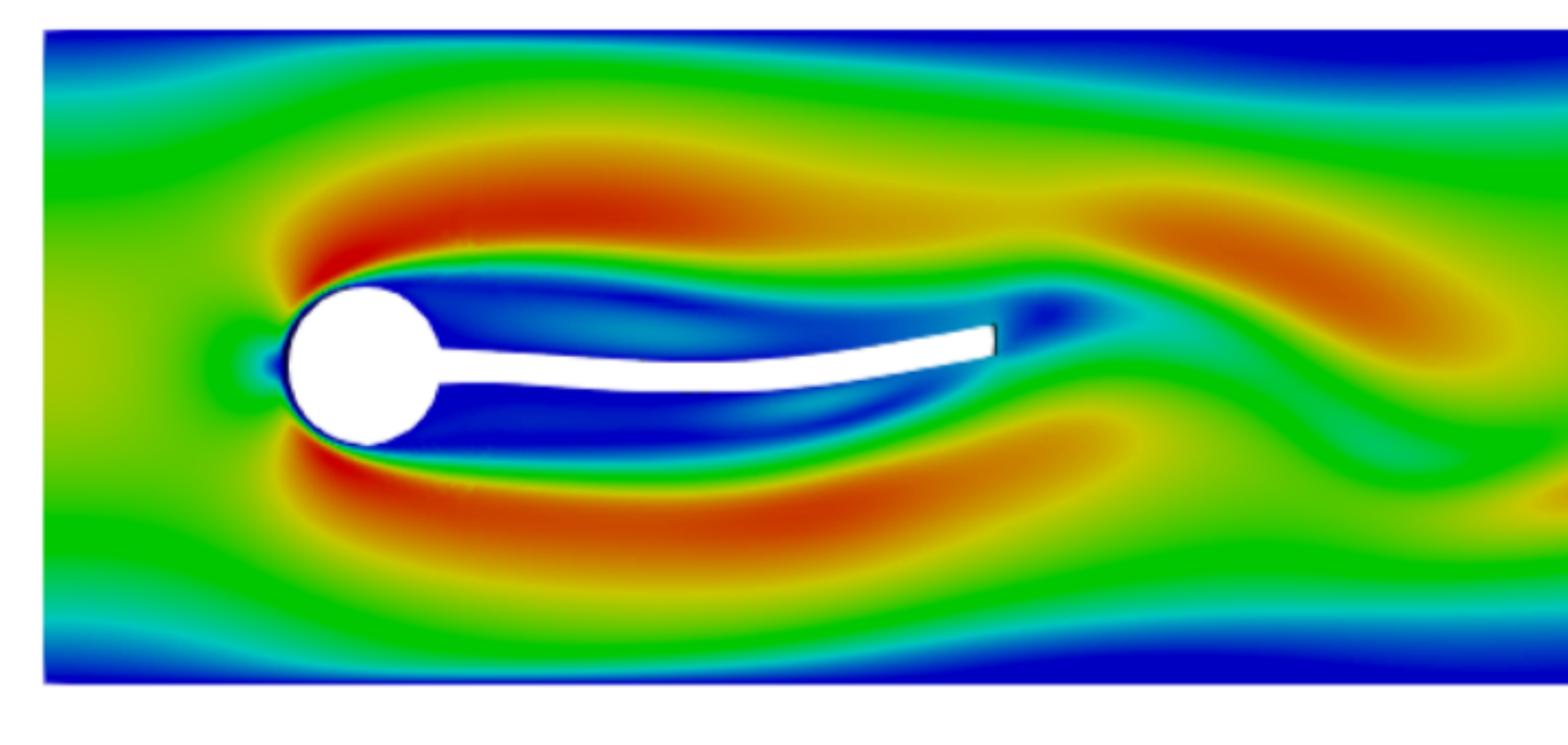}}}
\subfloat[simulation time = $12.5$~sec; \newline B= 2]{\scalebox{0.8}{\includegraphics[width=0.5\textwidth]{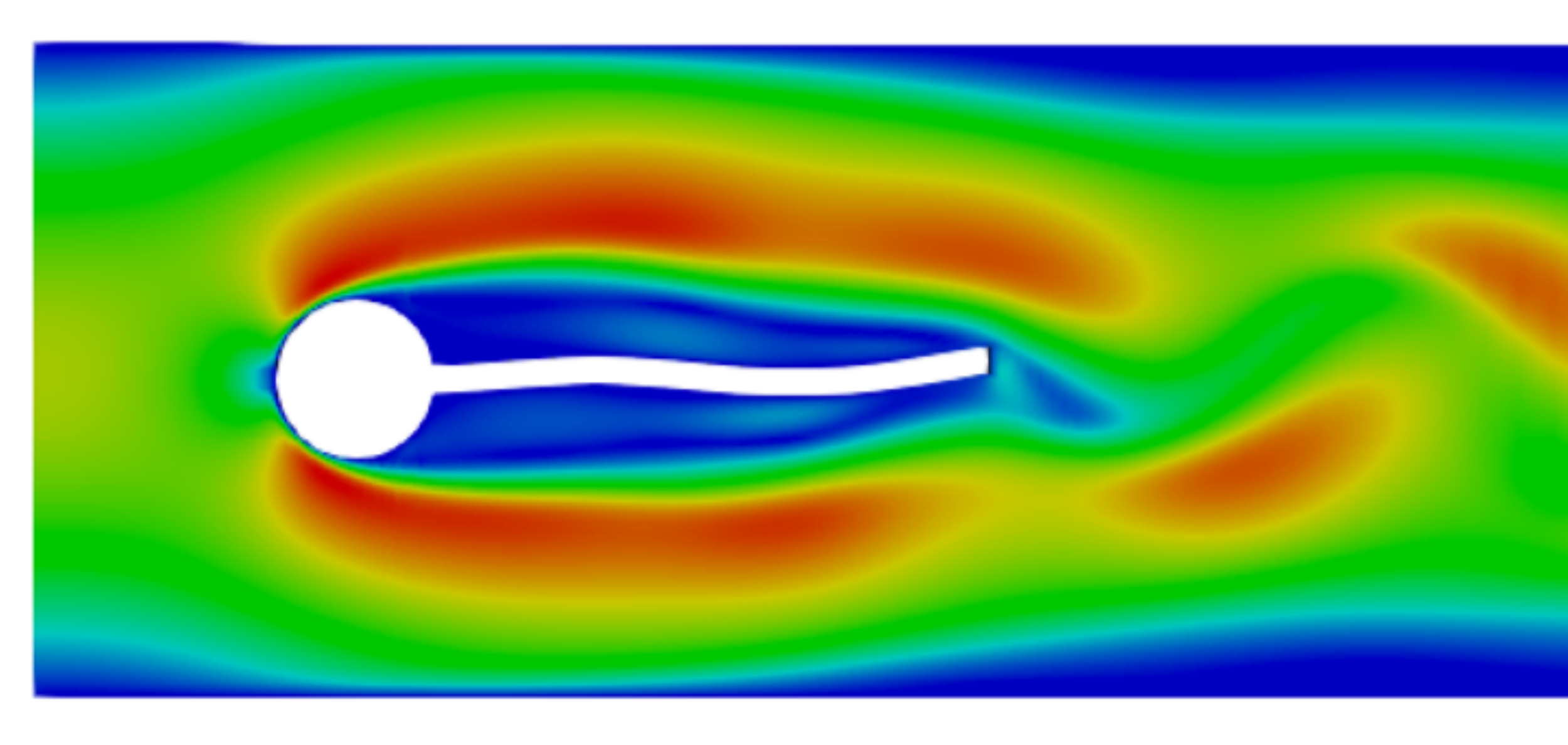}}}
\caption{Velocity profiles for simulations with beam stiffness $B=5.6$ and $B=2$ at time t = $6.25$~sec and $12.5$~sec,  respectively} \label{ex2_justi}
\end{figure}

\noindent This optimization problem may seem overly simple, because one might intuitively expect that larger stiffness values always yield a smaller tip displacement. However, the initialization data shows the opposite, as $B=2$ leads to a smaller $\delta$ than $B>3.1401$. To justify the results, velocity profiles of the test case are provided in Figure \ref{ex2_justi} where the beam deformation for constant elasticity functions, with $B=5.6$ and $B=2$, are compared. As shown in Figure \ref{ex2_justi}, under the same conditions, the cantilever with lower stiffness exhibits a higher mode of oscillation, under the action of the complex vortex shedding within the \textit{von K\'{a}rm\'{a}n vortex street}. As a result of this more complex motion, \textit{the less stiff beam exhibits a smaller tip deflection than the stiffer case.} This highlights some interesting, and counter intuitive, results that may arise within FSI contexts.\\

\begin{figure}[h!]
\captionsetup[subfigure]{justification=centering}
\centering
\subfloat[Tip displacement with uniform stiffness ($11^\text{th}$ iteration)\label{ex2iter}]{\scalebox{0.8}{\includegraphics[width=0.5\textwidth]{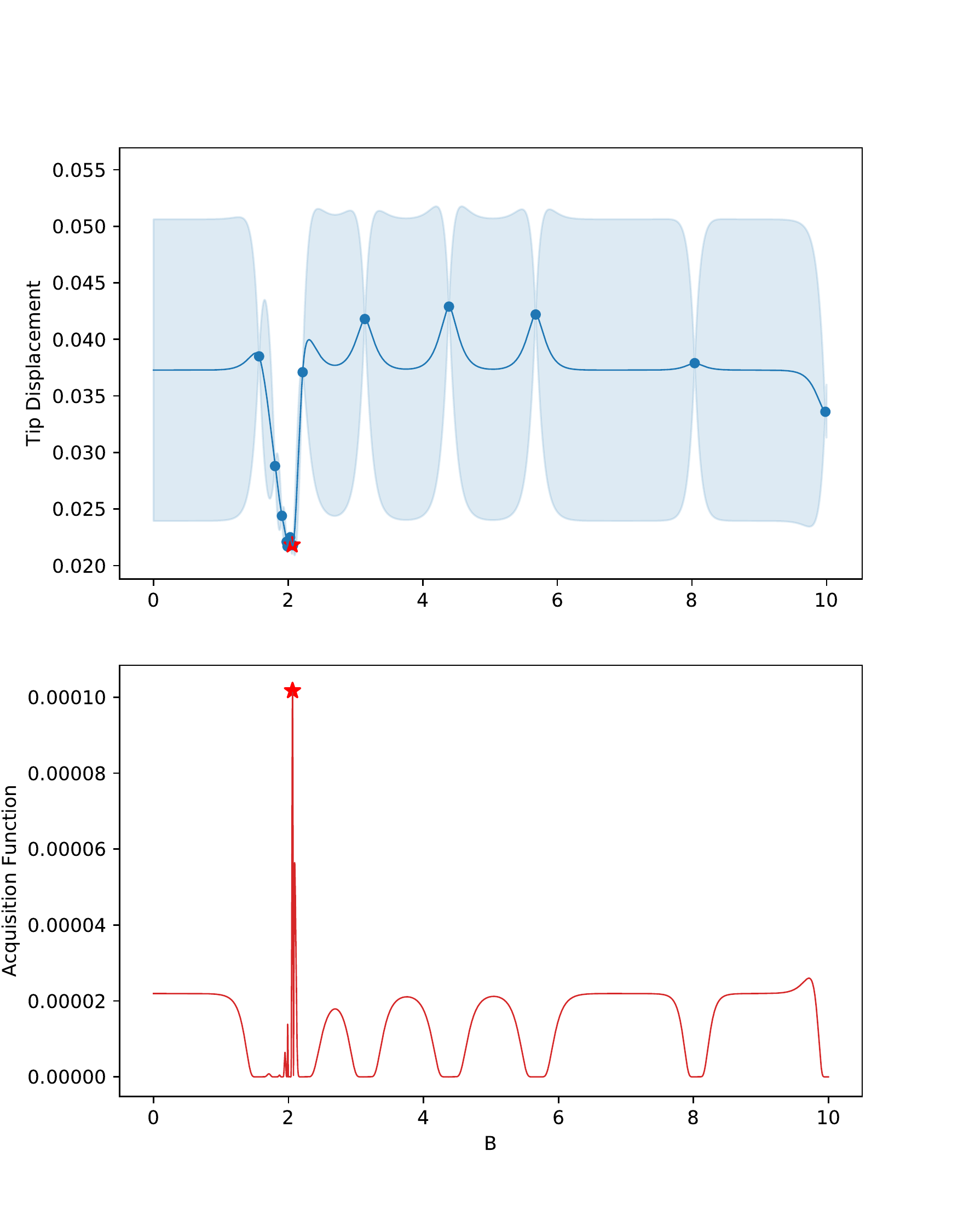}}}
\subfloat[Best minimum recorded at each BO iteration\label{ex2descent}] {\scalebox{0.8}{\includegraphics[width=0.5\textwidth]{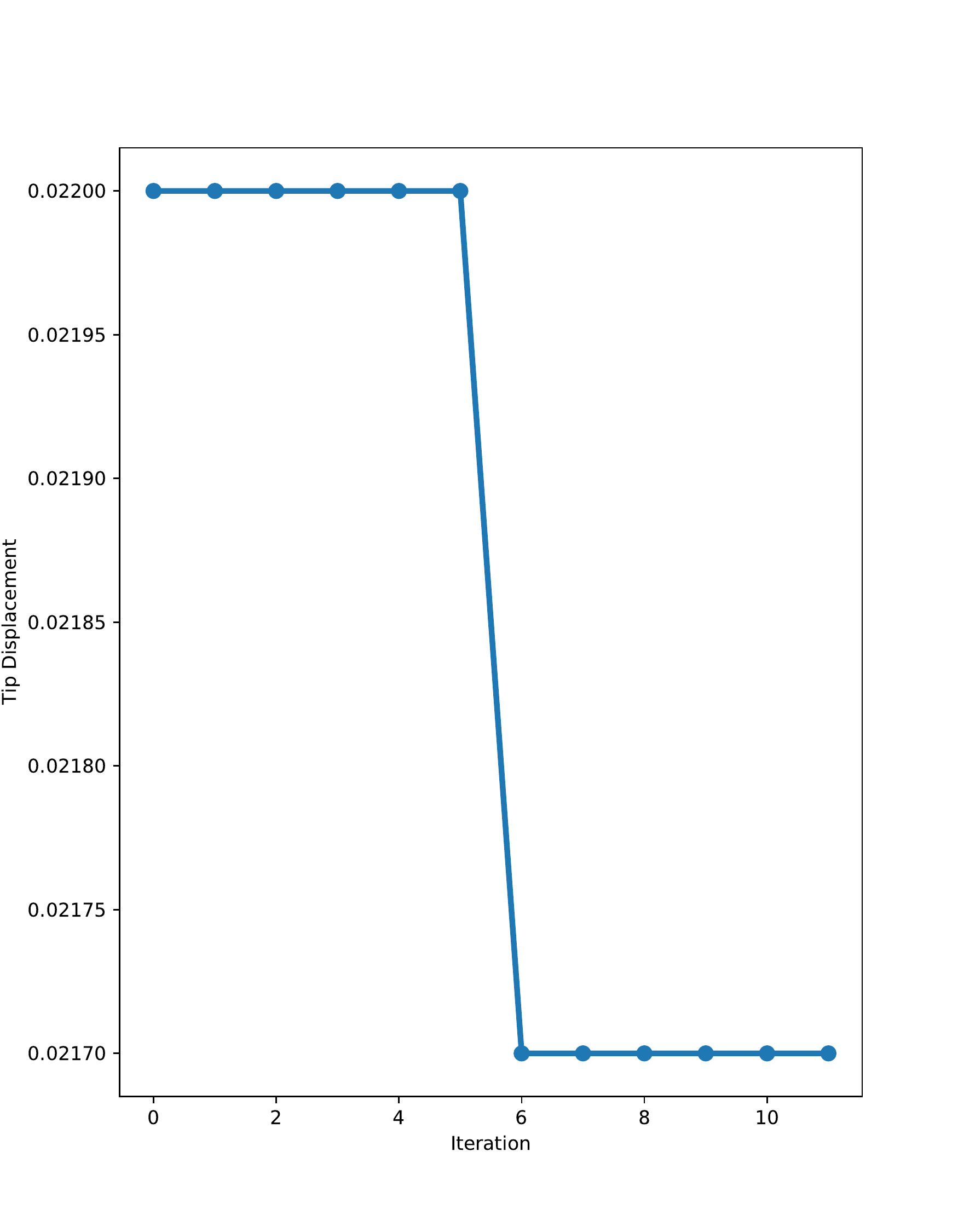}}}
\caption{Successive BO iterations - unconstrained uniform stiffness} \label{ex2}
\end{figure}

\noindent Figure \ref{ex2iter} shows the GP surrogate after 11 BO iterations and indicates that $\delta(B)$ follows a highly non-monotonic and non-convex behavior, with a significant drop in tip displacement for $B\approx2$. This drop is sharp, which makes the search for the global minimum difficult, but after 6 iterations, $(B_\text{min}=1.9907, \delta_\text{min}=0.0217)$ is found. This value is taken as the global minimum, since performing 5 additional iterations did not allow for further improvements (Figure \ref{ex2descent}).

\begin{table}[h]
\centering
\begin{tabular}{c|c|c}
Iteration \# & $B$ (MPa) & $\delta$ (m)\\\hline
Init & $8.0408$ & $0.0379$\\\hline
Init & $3.1401$ & $0.0418$\\\hline
Init & $5.6781$ & $0.0422$\\\hline
Init & $2.0000$ & $0.0220$\\\hline
1 & $1.5689$ & $0.0385$\\\hline
2 & $2.2160$ & $0.0371$\\\hline
3 & $1.9078$ & $0.0244$\\\hline
4 & $1.9774$ & $0.0221$\\\hline
5 & $2.0303$ & $0.0225$\\\hline
\textbf{6} & $\mathbf{1.9907}$ & $\mathbf{0.0217}$\\\hline
7 & $1.8059$ & $0.0288$\\\hline
8 & $9.9826$ & $0.0336$\\\hline
9 & $4.3898$ & $0.0429$\\\hline
10 & $2.0727$ & $0.0218$\\\hline
11 & $2.0599$ & $0.0220$\\
\end{tabular}
\caption{\label{ex2values} Successive BO iterations - unconstrained uniform stiffness}
\end{table}

\newpage
\subsection{Example 3: Unconstrained Non-Uniform Stiffness} \label{example3}

\begin{figure}[h!]
    \centering
    \includegraphics[width=0.5\textwidth]{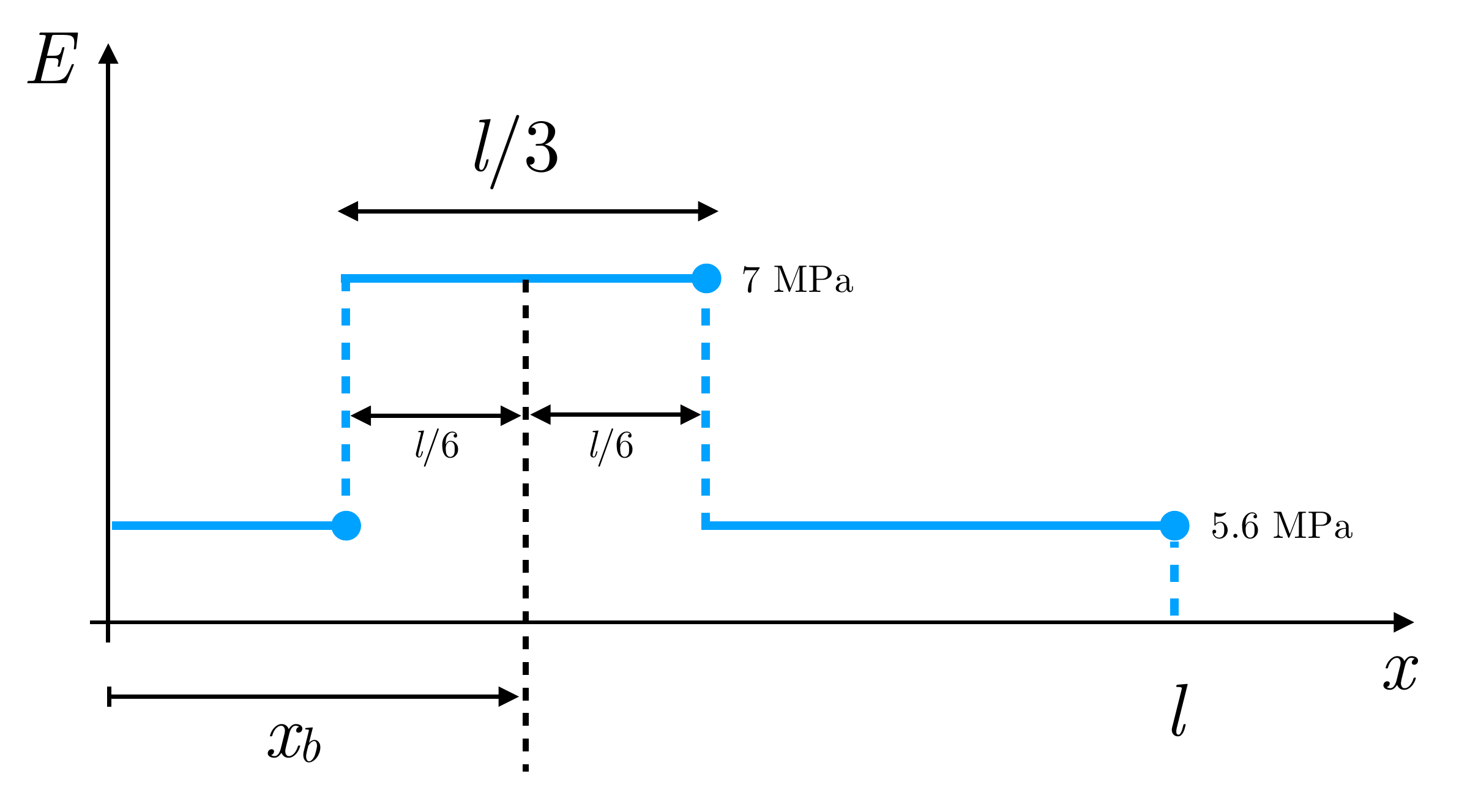}
    \caption{Stiffness along the beam given the box mid-point, $x_b$}
    \label{boxfunction}
\end{figure}

\noindent In this third example, we consider the stiffness function along the beam to be described by a \textit{box-function} as illustrated in Figure \ref{boxfunction}. Formally, the stiffness is defined by:
\begin{equation}
E(x)=
    \begin{cases}
    7.0\,\,\,\,\,\,\text{if}\,\,\,\,x\in[x_b-l/6,x_b+l/6]\\[5pt]
    5.6\,\,\,\,\,\,\text{else}
    \end{cases}
\end{equation}

\begin{figure}[h!]
\centering
\captionsetup[subfigure]{justification=centering}
\subfloat[Tip displacement with box function stiffness ($11^\text{th}$ iteration)\label{ex3iter}]{\scalebox{0.8}{\includegraphics[width=0.5\textwidth]{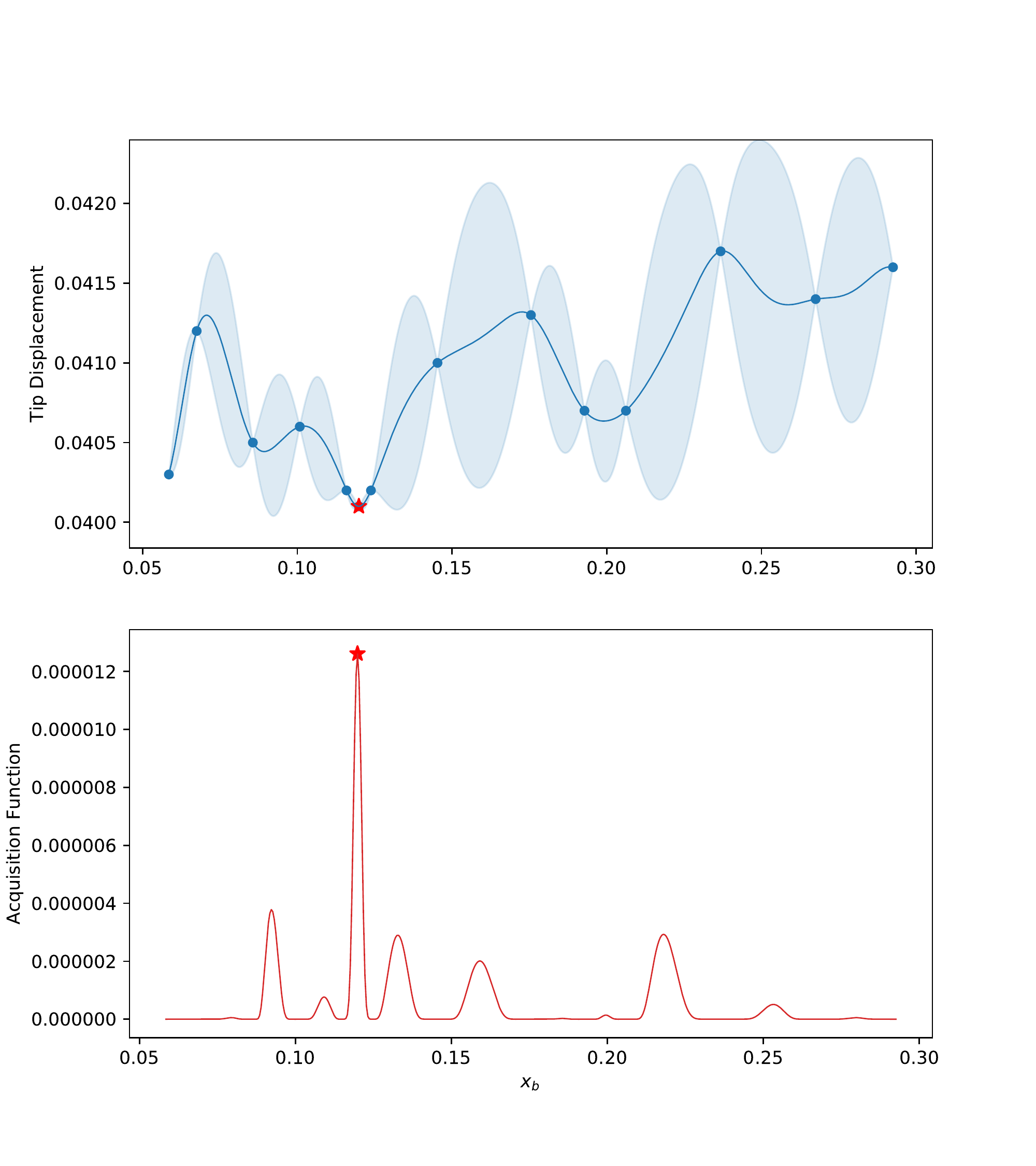}}}
\subfloat[Best minimum recorded at each BO iteration\label{ex3descent}] {\scalebox{0.8}{\includegraphics[width=0.5\textwidth]{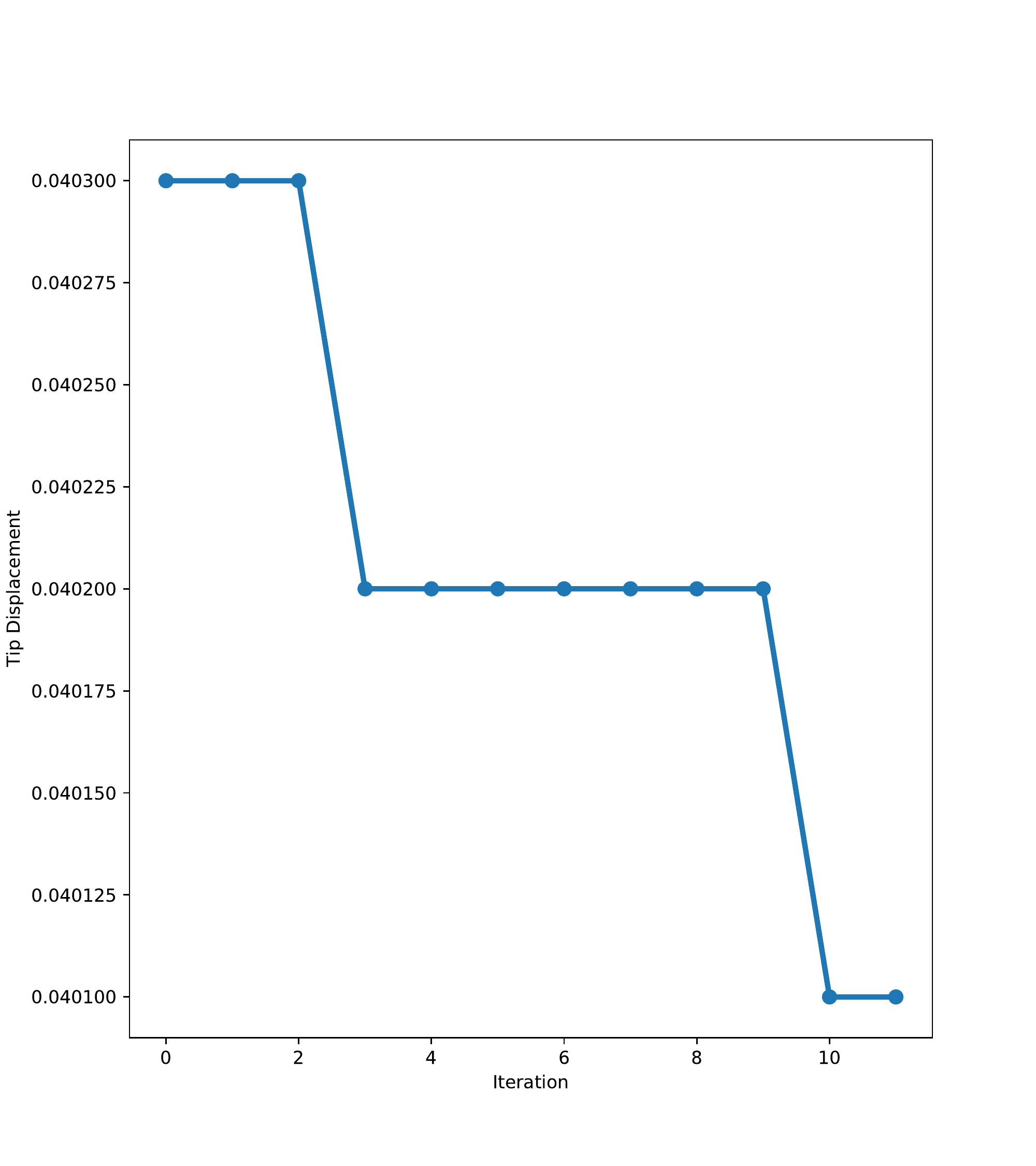}}}
\caption{Successive BO iterations - non-uniform stiffness} \label{ex3}
\end{figure}

\noindent Again, we are looking to minimize the vertical tip displacement, this time with respect to the box mid-point, $x_b$. Figure \ref{ex3iter} shows the surrogate GP after 11 iterations. The tip displacement tends to decrease, overall, when the zone of concentrated material stiffness (\textit{i.e} the \textit{box}) is positioned close to the beam's fixed end. Interestingly, the minimum tip deflection value, or $\delta_\text{min}=0.0401$, is not obtained for a box fully shifted to the far left (\textit{i.e.} corresponding to the fixed end), but rather for a box located at about a third of the beam length ($x_{b,\text{min}}=0.1199$). After 11 iterations, the next $x_b$ value proposed by the optimizer is $x_b=0.1199$, which has already been tried and yields the current best minimum. Therefore, it is unlikely to improve, and so the optimization is stopped there.

\begin{table}[h]
\centering
\begin{tabular}{c|c|c}
Iteration \# & $x_b$ (m) & $\delta$ (m)\\\hline
Init & $0.0585$ & $0.0403$\\\hline
Init & $0.1755$ & $0.0413$\\\hline
Init & $0.2925$ & $0.0416$\\\hline
1 & $0.0856$ & $0.0405$\\\hline
2 & $0.0675$ & $0.0412$\\\hline
3 & $0.1238$ & $0.0402$\\\hline
4 & $0.2368$ & $0.0417$\\\hline
5 & $0.2062$ & $0.0407$\\\hline
6 & $0.1159$ & $0.0402$\\\hline
7 & $0.1008$ & $0.0406$\\\hline
8 & $0.2675$ & $0.0414$\\\hline
9 & $0.1453$ & $0.0410$\\\hline
\textbf{10} & $\mathbf{0.1199}$ & $\mathbf{0.0401}$\\\hline
11 & $0.1928$ & $0.0407$\\
\end{tabular}
\caption{\label{ex3values} Successive BO iterations - non-uniform stiffness}
\end{table}

\newpage
\section{Composite Material Properties Selection for UUAV Sail Plane} \label{UUAV}
\noindent Fortified with our experience with verification and bridging simulations, we now turn our attention towards a realistic application: a UUAV sail plane made from FRP (fiber reinforced polymer), as shown in Figure \ref{sailplane_geo}. The sail plane geometry is fixed: $0.4$~m long and $0.01$~m wide, with some adjustable angle of attack, $\theta$ (with respect to the horizontal). The leading edge of the sail plane is $0.038$~m thick; reducing linearly with length, to assume a value of $0.004$~m thickness at the trailing edge. The sail plane has a density of $2000$~kg/m$^3$. A uniform inflow velocity of $5$~m/s is prescribed along the left boundary. We wish to find the optimal (uniform) stiffness, $E$, and angle of attack $\theta$, that minimizes the drag over time, while imposing conditions on the lift, the vertical tip displacement, as well as the outlet pressure differential. Unlike previous cases, this is a 2-parameters optimization problem:

\begin{figure}[h!]
\centering
\includegraphics[width=1\textwidth]{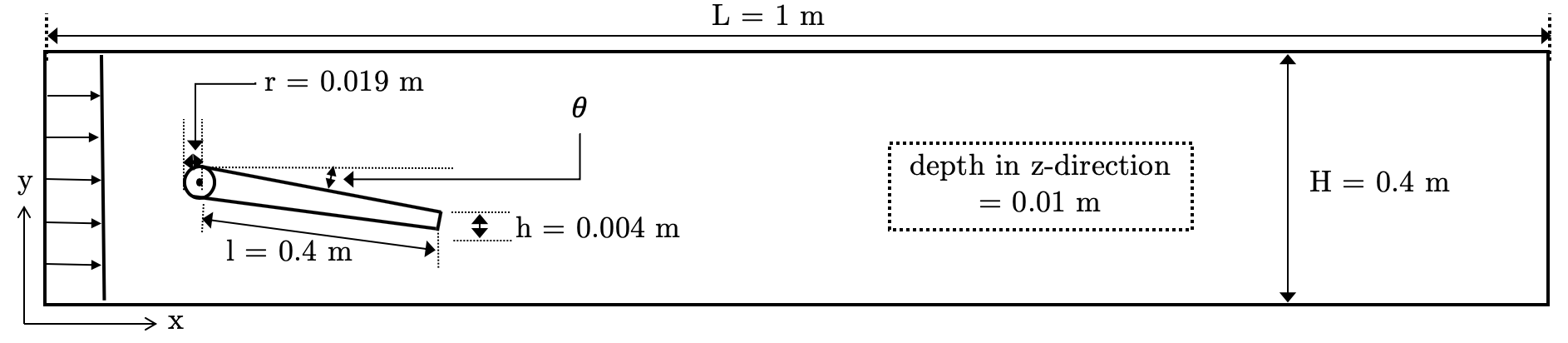}
\caption{UUAV geometry configuration}\label{sailplane_geo}
\end{figure}

\begin{equation}
\text{Find}\,\,\,\,\,\min_{\theta,E}F_D\,\,\,\,\,\text{s.t.}\,\,\,\,\,\begin{cases}
F_L\geq F_{L,c}\\
\delta\leq \delta_c\\
\Delta p\leq \Delta p_c
\end{cases}
\end{equation}

\begin{figure}[h!]
    \centering
    \includegraphics[width=0.5\textwidth]{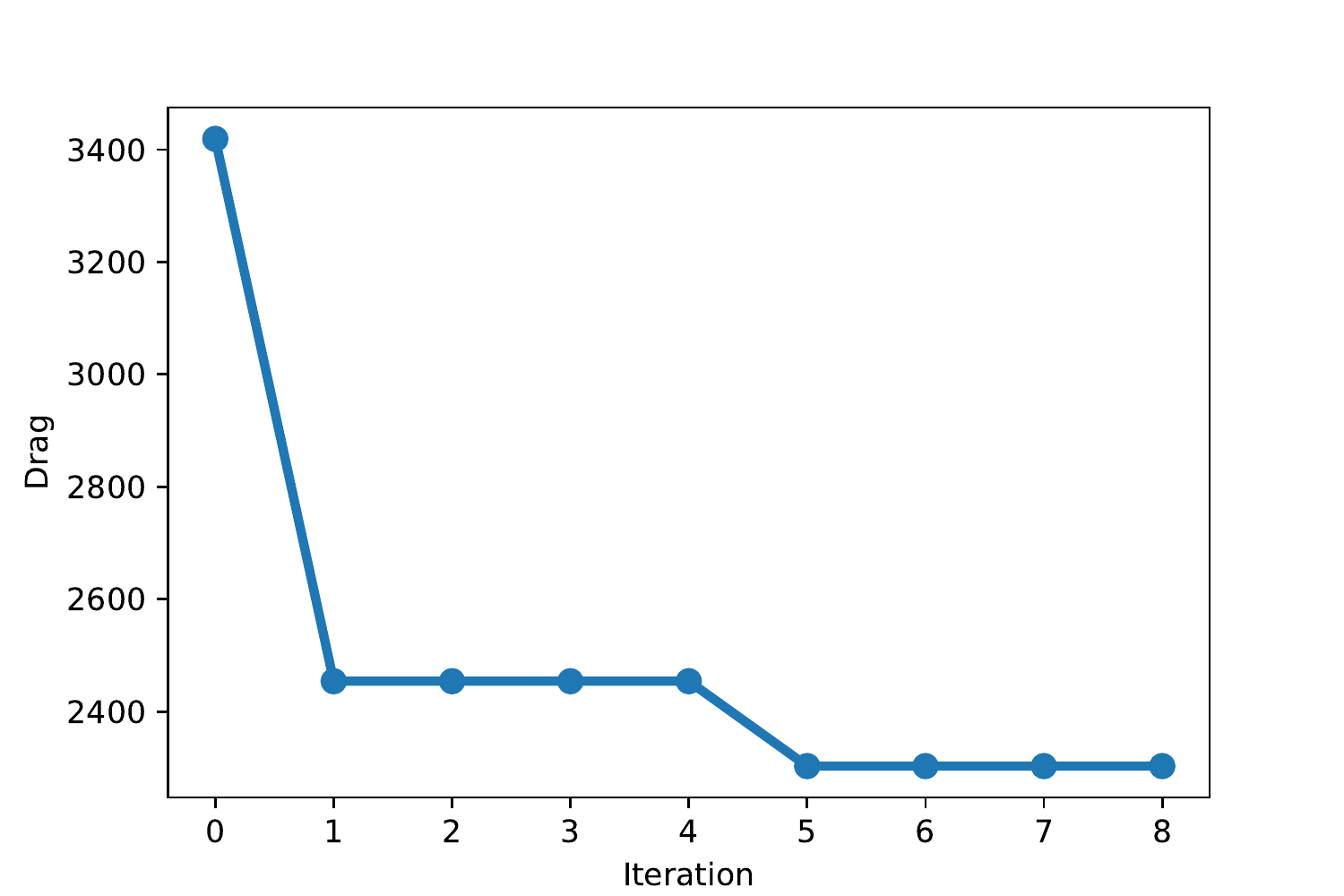}
    \caption{Best minimum recorded at each BO iteration - UUAV sail plane}
    \label{ex4descent}
\end{figure}

\begin{figure}[h!]
\centering
\captionsetup[subfigure]{justification=centering}
\subfloat[Initialization\label{ex4iter0}]{\scalebox{0.925}{\includegraphics[width=0.33\textwidth, height=3cm]{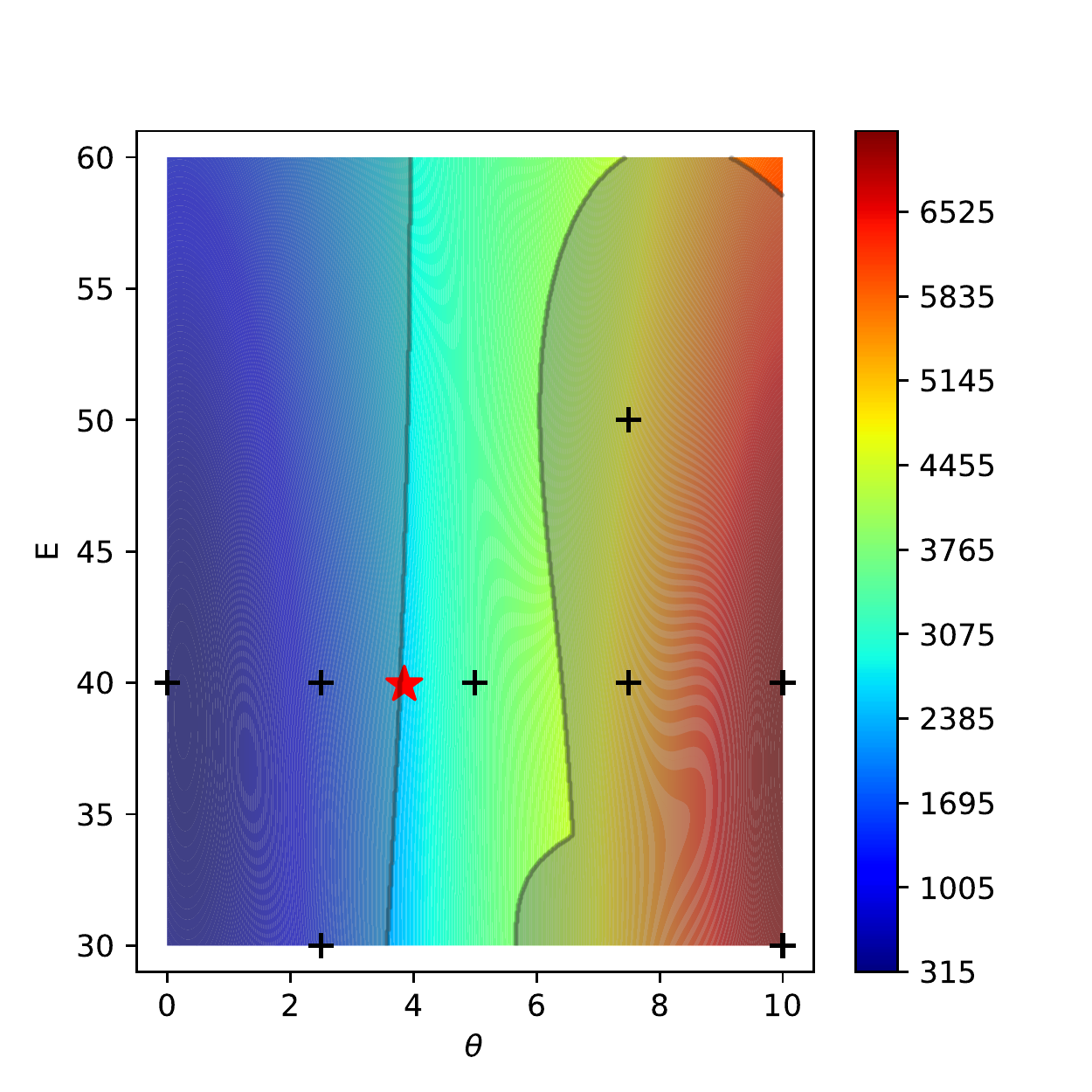}}}
\subfloat[Iteration 1\label{ex4iter1}] {\scalebox{0.925}{\includegraphics[width=0.33\textwidth, height=3cm]{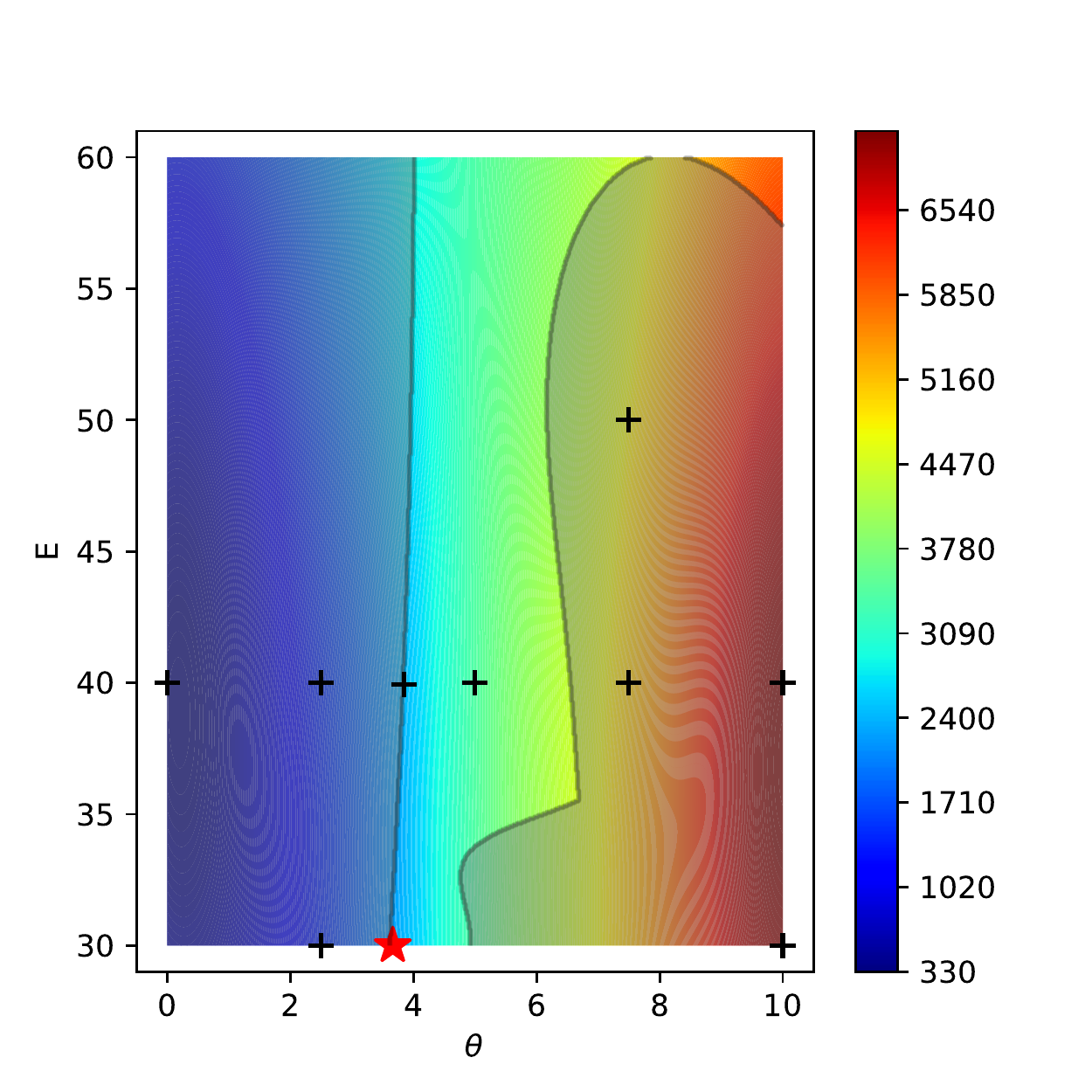}}}
\subfloat[Iteration 2\label{ex4iter2}]{\scalebox{0.925}{\includegraphics[width=0.33\textwidth, height=3cm]{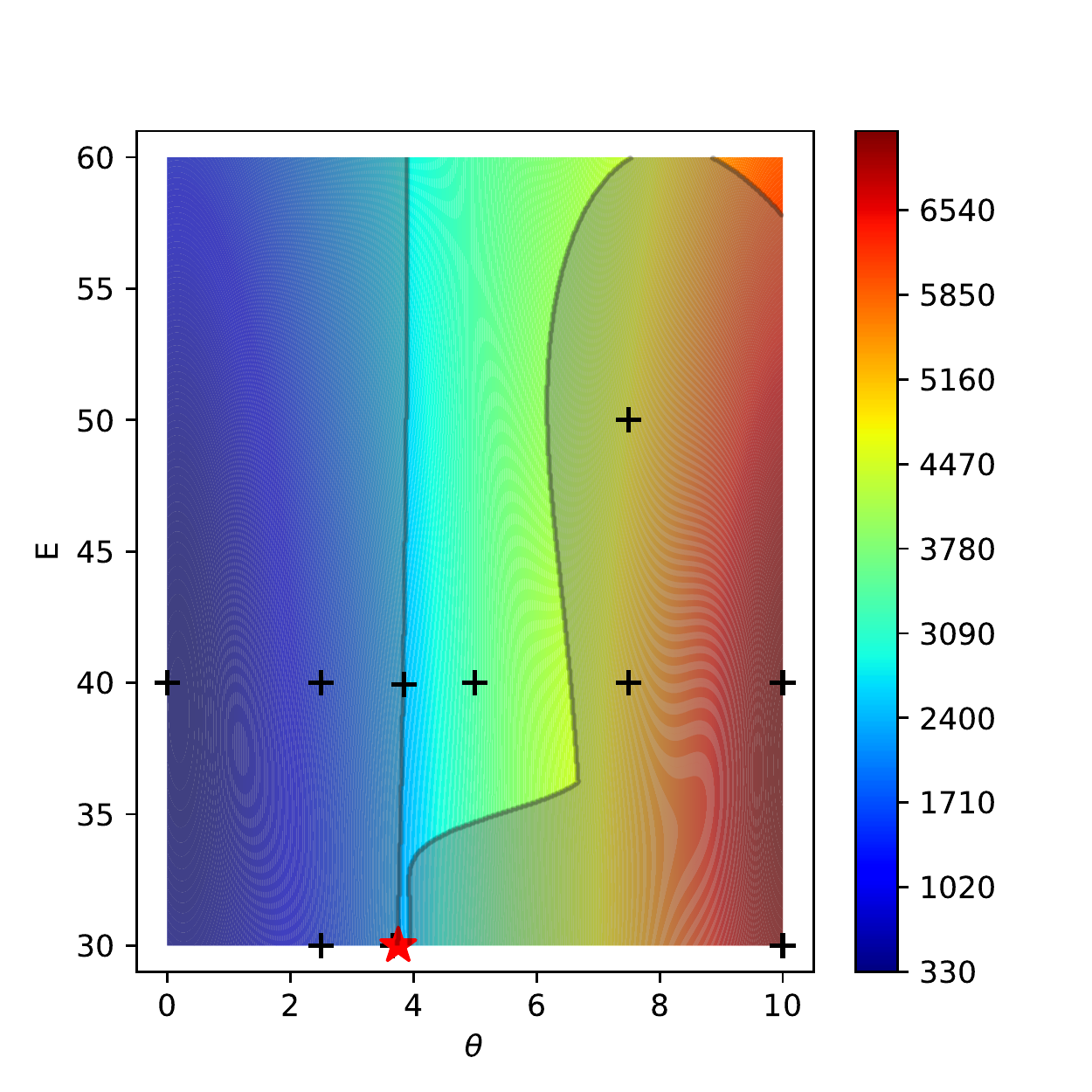}}}\\
\subfloat[Iteration 3\label{ex4iter3}]{\scalebox{0.925}{\includegraphics[width=0.33\textwidth, height=3cm]{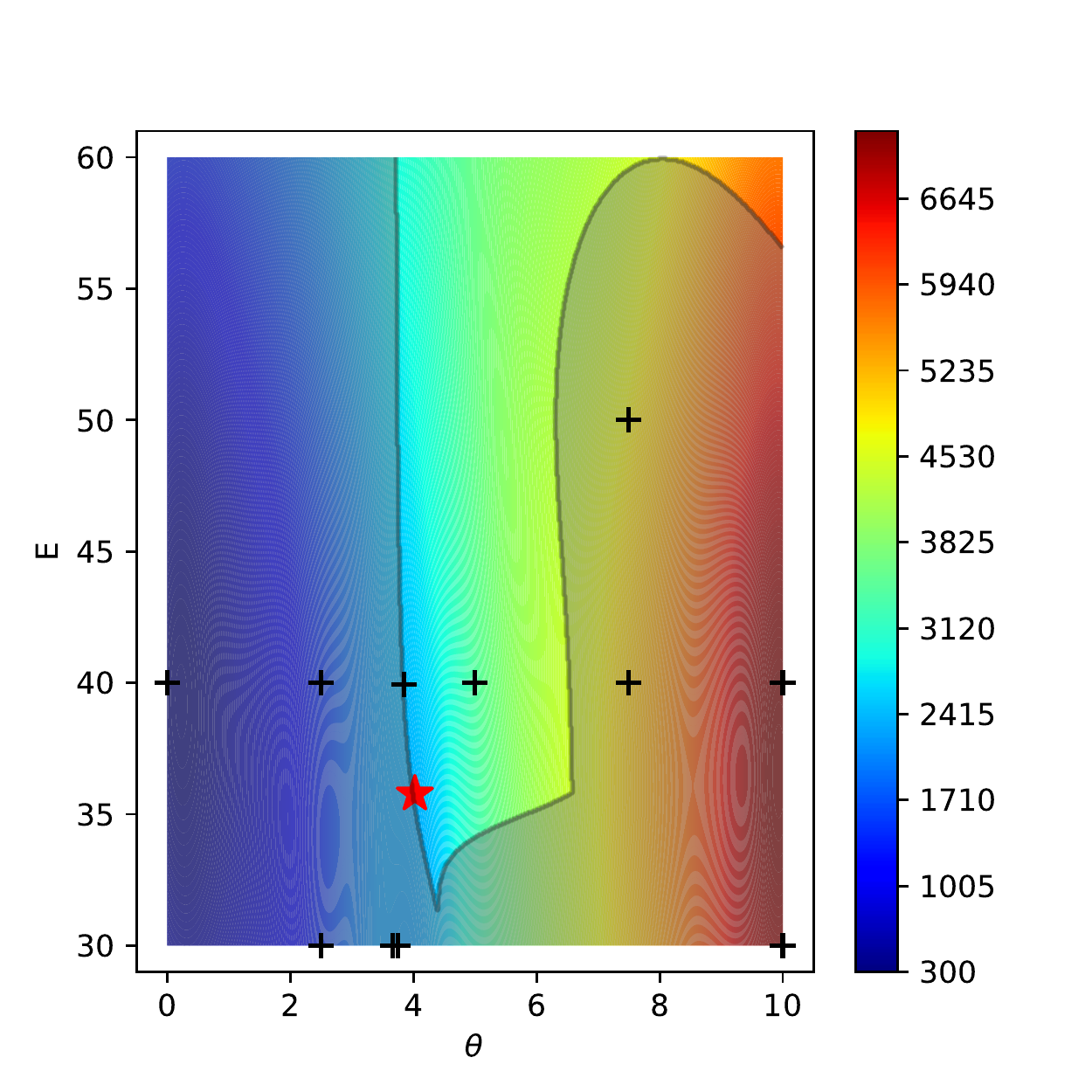}}}
\subfloat[Iteration 4\label{ex4iter4}]{\scalebox{0.925}{\includegraphics[width=0.33\textwidth, height=3cm]{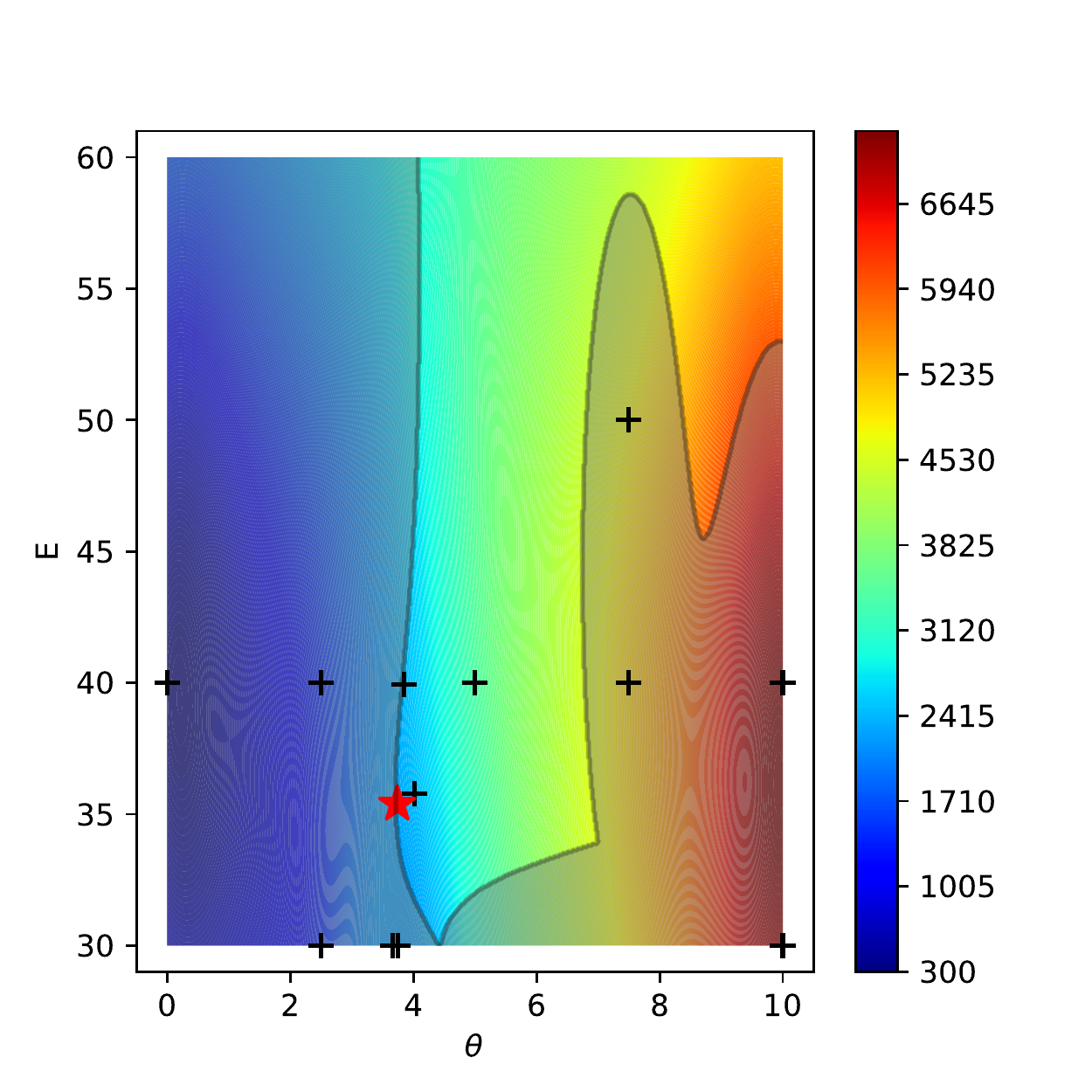}}}
\subfloat[Iteration 5\label{ex4iter5}]{\scalebox{0.925}{\includegraphics[width=0.33\textwidth, height=3cm]{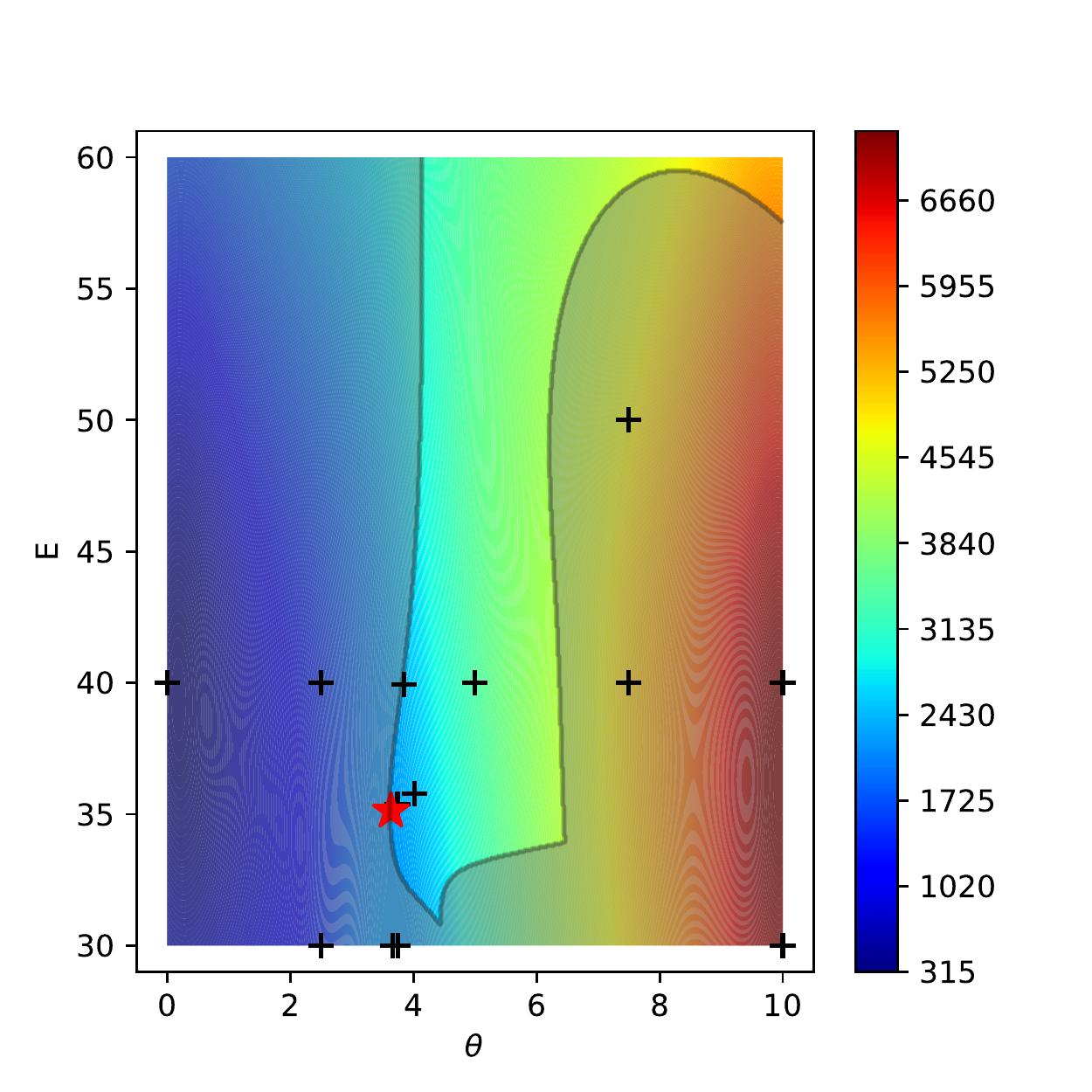}}}\\
\subfloat[Iteration 6\label{ex4iter6}]{\scalebox{0.925}{\includegraphics[width=0.33\textwidth, height=3cm]{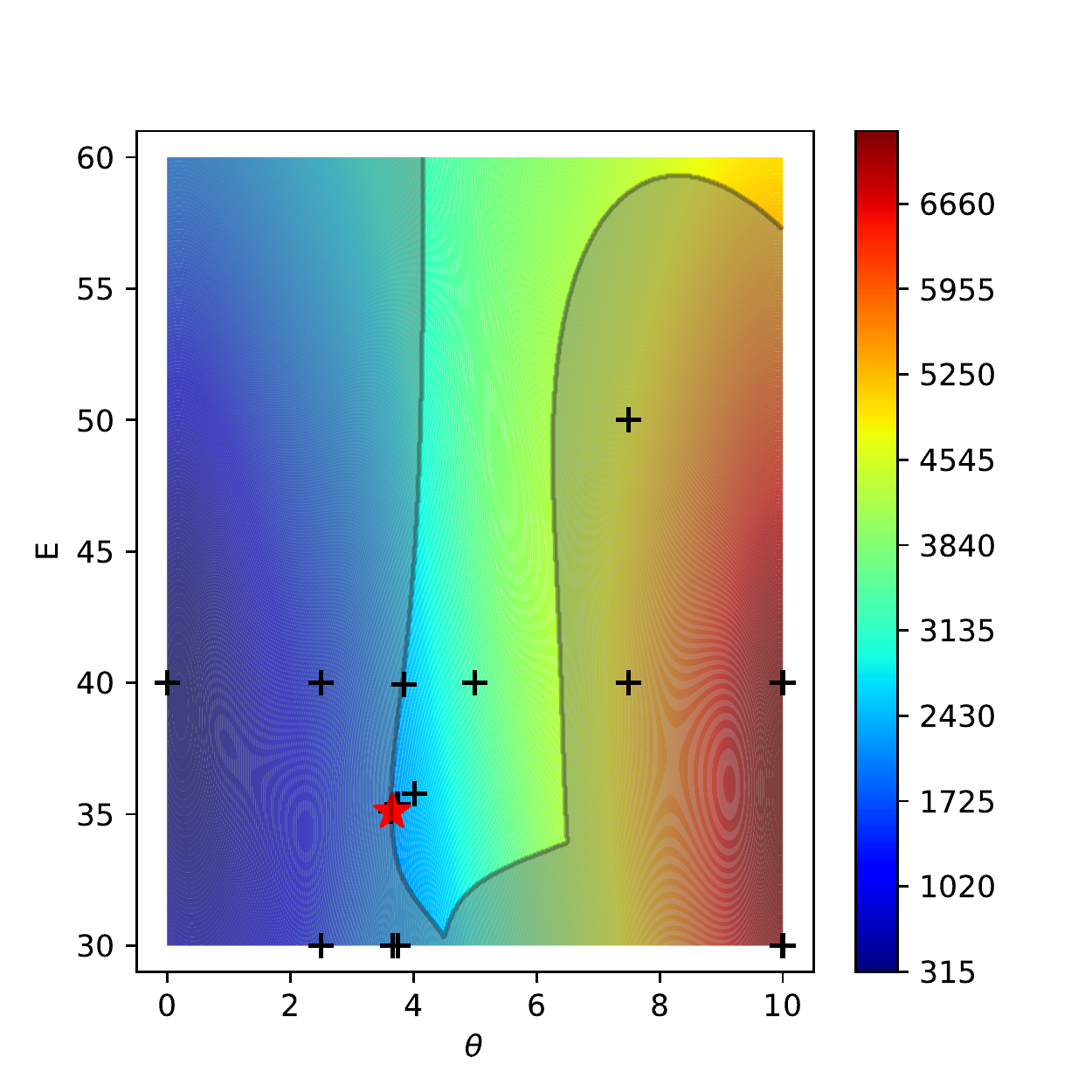}}}
\subfloat[Iteration 7\label{ex4iter7}]{\scalebox{0.925}{\includegraphics[width=0.33\textwidth, height=3cm]{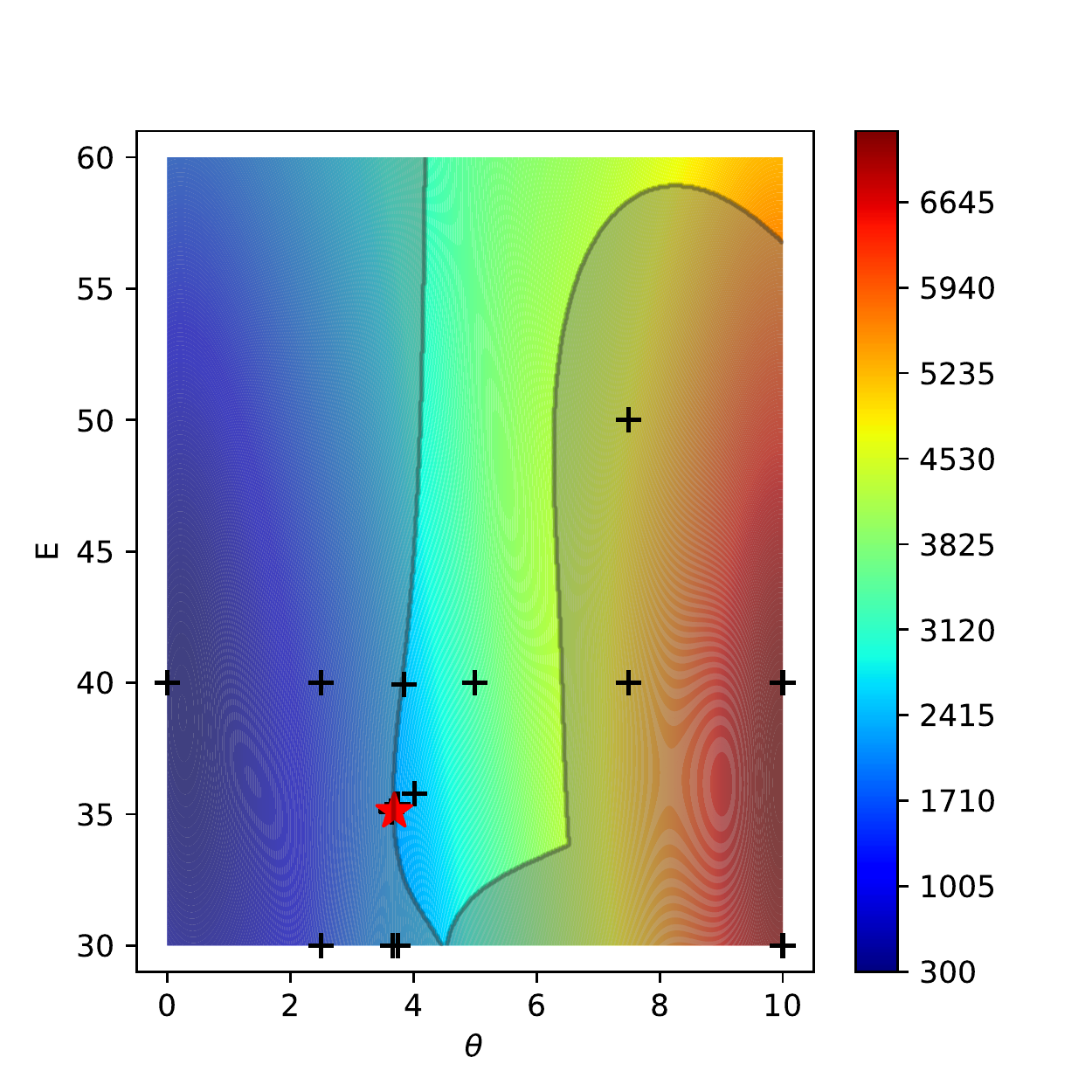}}}
\subfloat[Iteration 8\label{ex4iter8}]{\scalebox{0.925}{\includegraphics[width=0.33\textwidth, height=3cm]{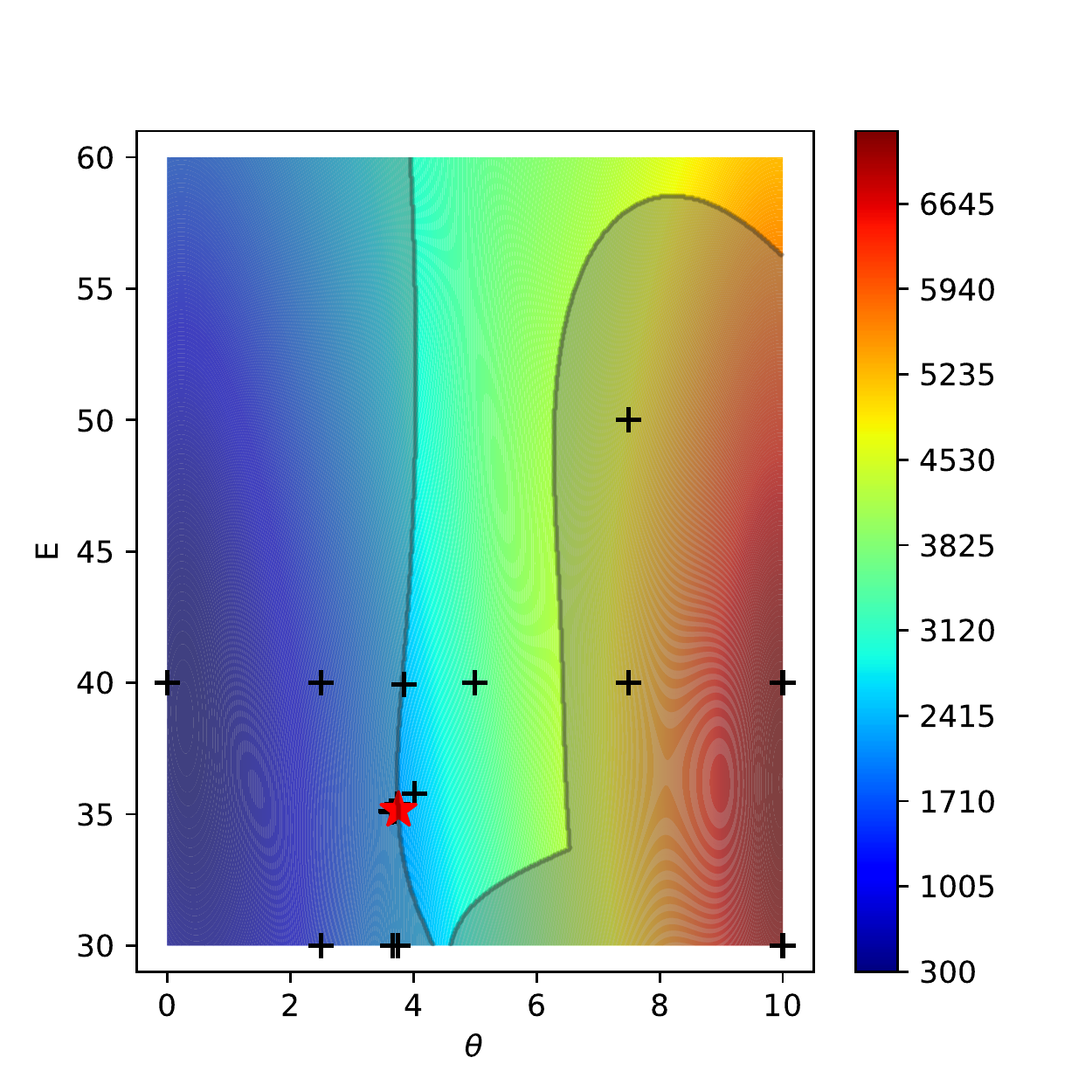}}}
\caption{Successive BO iterations - UUAV sail plane. The color surface plots represent the GP predictive mean for the drag, and the shaded dark areas represent the unfeasible input regions imposed on the surrogate GPs for the lift, tip displacement, and pressure.} \label{ex4iter}
\end{figure}

\noindent The drag $F_D$ and lift $F_L$ values are respectively taken as the sum of the total pressure acting on the sail plane in the tangential and normal direction, with respect to the orientation of the sail plane, over the time span $[0.5~\text{sec},3.5~\text{sec}]$. $\delta$ is the plane vertical tip displacement. The pressure differential, $\Delta p$, is taken as $\Delta p=p_\text{max}-p_\text{min}$, where $p_\text{max}$ and $p_\text{min}$ are the pressures recorded at the outlet, at a height of $5/8H$, during the time interval $[0.5~\text{sec},3.5~\text{sec}]$.  The constraints are $F_{L,c}=12000$ N, $\delta_c=2.5$ mm, and $\Delta p_c=120$ kPa. The Bayesian Optimizer is initialized with eight data points (since the input space is now in two dimensions, more data are required for fitting the surrogate GPs properly). Table \ref{ex4values} shows the successive input parameters proposed by the optimizer, along with the corresponding FSI analysis results. Figure \ref{ex4iter} shows the predictive mean of the drag force predicted by the surrogate GP at, each BO iteration, with the feasible/unfeasible inputs regions (given the constraint set for the surrogate GPs). Figure \ref{ex4descent} shows the best minimum found after each iteration. The drag appears to follow a very smooth and fairly monotonic behavior, and the optimizer converges quickly toward a minimum. As seen in Figure \ref{ex4iter}, the feasible regions provide significant freedom for $E$ to vary, whereas the feasible regions for $\theta$ are much more narrow. This is intuitive since one might expect the angle of the sail plane to influence the drag more than the stiffness. Yet, the optimal stiffness value is not trivial. Given the optimizer progress after 8 iterations (Figure \ref{ex4descent}), it seems unlikely that a significantly better minimum drag value can be achieved, and the iteration loop is stop there. The global minimum is taken as ($\theta_\text{min}=3.74,\,E_\text{min}=35.39,\,F_\text{D,min}=2303$).

\begin{table}[h]
\centering
\begin{tabular}{c|c|c|c|c|c|c}
Iteration \# & $\theta$ ($^{\circ}$) & $E$ (MPa)& $F_D$ (N) & $F_L$ (N) & $\delta$ (mm)& $\Delta p$ (kPa)\\ \hline
Init & $0.00$ & $40.00$ & $ 350$ & $ 3540$ & $0.565$ & $9.50$\\\hline
Init & $2.50$ & $40.00$ & $1482$ & $ 9214$ & $1.206$ & $39.90$\\\hline
Init & $5.00$ & $40.00$ & $3419$ & $13736$ & $2.146$ & $89.70$\\\hline
Init & $7.50$ & $40.00$ & $5042$ & $16183$ & $2.593$ & $141.90$\\\hline
Init & $10.00$ & $40.00$ & $7148$ & $18333$ & $3.399$ & $165.60$\\\hline
Init & $2.50$ & $30.00$ & $1452$ & $10051$ & $1.813$ & $42.40$\\\hline
Init & $7.50$ & $50.00$ & $4767$ & $16336$ & $1.884$ & $139.60$\\\hline
Init & $10.00$ & $30.00$ & $7026$ & $17716$ & $4.840$ & $173.80$\\\hline
 1 & $3.85$ & $39.93$ & $2454$ & $12025$ & $1.739$ & $74.53$\\\hline
 2 & $3.66$ & $30.00$ & $2256$ & $11891$ & $2.355$ & $61.83$\\\hline
 3 & $3.75$ & $30.00$ & $2247$ & $11320$ & $2.183$ & $71.42$\\\hline
 4 & $4.02$ & $35.77$ & $2466$ & $12467$ & $1.749$ & $89.06$\\\hline
 \textbf{5} & $\mathbf{3.74}$ & $\mathbf{35.39}$ & $\mathbf{2303}$ & $\mathbf{12184}$ & $\mathbf{1.956}$ & $\mathbf{50.37}$\\\hline
 6 & $3.63$ & $35.12$ & $2172$ & $11961$ & $1.939$ & $77.97$\\\hline
 7 & $3.65$ & $35.10$ & $2285$ & $11858$ & $1.941$ & $76.85$\\\hline
 8 & $3.69$ & $35.11$ & $2164$ & $11547$ & $1.934$ & $78.80$\\
\end{tabular}
\caption{\label{ex4values} Successive BO iterations - UUAV sail plane}
\end{table}

\newpage
\section{Conclusion} \label{conclusion}
\noindent This paper has demonstrated a principled approach to design using high fidelity FSI simulations. Our proposed approach consists of three stages. Firstly, verify and/or validate the FSI computational tool using canonical cases and/or experiment results from the literature with design configurations that are close to the ultimate application. This helps identify the optimal spatiotemporal discretization, as well as numerical schemes, for the design application. Subsequently, incrementally and systematically increase the complexity of the verification case to bring it closer to the design context, using what we term \textit{bridge simulations}. This step provides information to gauge whether adjustments need to be made to further improve the computational model. Finally, apply the verified FSI framework, thus now trusted, to the design problem of interest. \\
\newline
Our experience with BO has shown it to be able to handle complicated design scenarios that have multiple design parameters, and multiple constraints. Furthermore, it is able to effectively optimize design cases that exhibit both smooth and non-smooth objective functions, even with counter-intuitive optimal parameters directions.

\section{Acknowledgement} \label{Acknowledgement}
\noindent The support of the Office of Naval Research (ONR), under the grant N00014-19-1-2034 and contract N6833518C0217, is gratefully acknowledged.

\section*{References}

\bibliography{references}

\end{document}